%% file: paper.tex
\documentclass[]{joyenfra}

\usepackage[toc,page,header]{appendix}

\usepackage{minitoc}
\usepackage{cleveref} 
\usepackage{subcaption}
\usepackage{booktabs}
\usepackage{graphicx}
\usepackage{subcaption}

\usepackage[table]{xcolor}
\usepackage{makecell}
\usepackage{multirow}
\usepackage{booktabs}

\newcolumntype{L}[1]{>{\raggedright\arraybackslash}p{#1}}
\newcolumntype{C}[1]{>{\centering\arraybackslash}p{#1}}
\newcolumntype{Y}{>{\raggedright\arraybackslash}X}

\usepackage{fancyhdr} 

\input{macro}
\title{\includegraphics[height=1em]{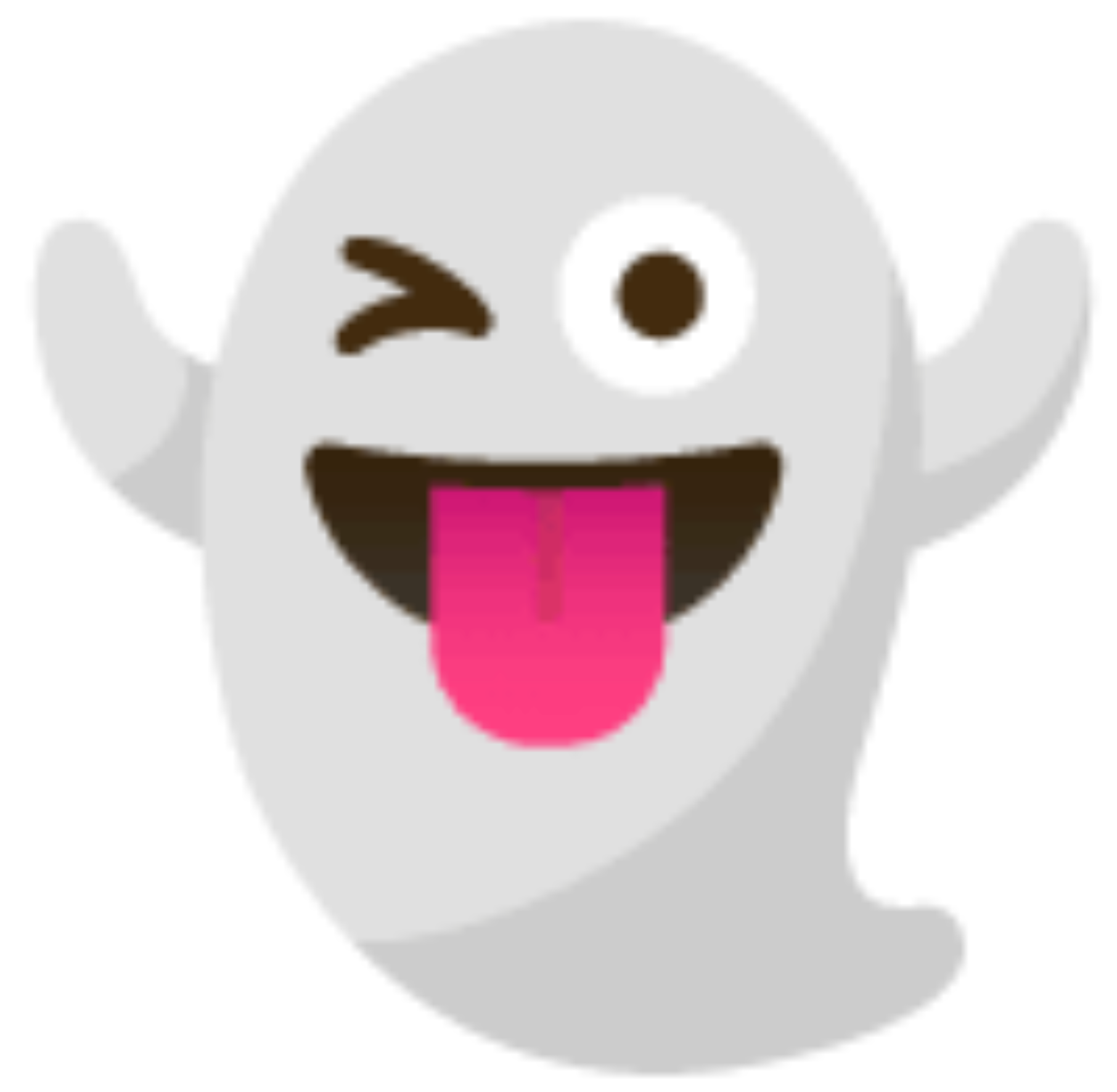}SPECTRE: Hybrid Ordinary-Parallel Speculative Serving for Resource-Efficient LLM Inference}

\affiliation[1]{Jincheng Xie\quad $^2$Yawen Ling\quad $^2$Qi Xiao\\ $^2$Feiyu Zhang\quad $^1$Zhongyi Huang\quad $^2$Wen Hu$^\star$ \quad $^3$Yu Zheng$^\star$}
\affiliation[1]{Tsinghua  University \quad $^2$AI Infra Team at JDT \\ $^3$JD iCity, JD Technology, JD Intelligent Cities Research}
\affiliation[]{xiejc22@mails.tsinghua.edu.cn\quad \{lingyawen1, xiaoqi.31, 
zhangfeiyu.17\}@jd.com\quad zhongyih@tsinghua.edu.cn\quad msyuzheng@outlook.com\quad huwen.31@jd.com}

\affiliation[*]{Corresponding Author}

\input{tex/00_abstract}

\date{\today}

\begin{document}

\maketitle

\input{tex/01_intro}
\input{tex/03_analysis}
\input{tex/04_method}
\input{tex/05_experiment}
\input{tex/06_conclusion}


\bibliographystyle{unsrtnat}
\bibliography{main}

\input{tex/A0_implementation_details}
\input{tex/A_analysis}
\input{tex/A1_more_experiment}
\input{tex/A2_experiment_detail}

\clearpage



\end{document}

%% file: macro.tex
\usepackage{natbib}
\usepackage{latexsym}

\usepackage{url}
\usepackage{amssymb}
\usepackage[utf8]{inputenc}
\usepackage{microtype}
\usepackage{booktabs}
\usepackage{pifont} 
\usepackage{multirow}
\usepackage{makecell}
\usepackage{paralist}
\usepackage{xspace}
\usepackage{color}
\usepackage{xcolor}
\usepackage{colortbl}
\usepackage{adjustbox}
\usepackage{hyperref} 
\usepackage[edges]{forest}
\usepackage{tikz} 
\usepackage{caption}
\usepackage{amsfonts}

\hypersetup{
    colorlinks,
    linkcolor={blue!80!black},
    citecolor={blue!80!black},
}
\tikzset{
    root/.style =             {align=center, text width=1cm, rounded corners=3pt, line width=0.3mm, fill=gray!10, draw=gray!80, font=\small},
    demographic/.style =         {align=center, text width=1.8cm, rounded corners=3pt, line width=0.3mm, fill=blue!10, draw=blue!80, font=\footnotesize},
    demographic_work/.style =    {align=center, text width=10cm, rounded corners=3pt, line width=0.3mm, fill=blue!10, draw=blue!0, font=\footnotesize},
    character/.style =         {align=center, text width=1.8cm, rounded corners=3pt, line width=0.3mm, fill=red!10, draw=red!80, font=\footnotesize},
    character_work/.style =    {align=center, text width=10cm, rounded corners=3pt, line width=0.3mm, fill=red!10, draw=red!0, font=\footnotesize},
    personalization/.style =           {align=center, text width=1.8cm, rounded corners=3pt, line width=0.3mm, fill=cyan!10, draw=cyan!80, font=\footnotesize},
    personalization_work/.style =      {align=center, text width=10cm, rounded corners=3pt, line width=0.3mm, fill=cyan!10, draw=cyan!0, font=\footnotesize},
    risk/.style =         {align=center, text width=1.8cm, rounded corners=3pt, line width=0.3mm, fill=orange!10, draw=orange!80, font=\footnotesize},
    risk_work/.style =    {align=center, text width=10cm, rounded corners=3pt, line width=0.3mm, fill=orange!10, draw=orange!0, font=\footnotesize},
}

\usepackage{CJK}

%% file: tex/00_abstract.tex
\abstract{
LLM serving platforms are increasingly deployed as multi-model cloud systems, where user demand is often long-tailed: a few popular large models receive most requests, while many smaller tail models remain underutilized. We propose \textbf{SPECTRE} (Parallel \textbf{SPEC}ulative Decoding with a Multi-\textbf{T}enant \textbf{RE}mote Drafter), a serving framework that reuses underutilized tail-model services as remote drafters for heavily loaded large-model services through speculative decoding. SPECTRE enables draft generation and target-side verification to run in parallel, and makes such parallelism effective through three techniques: a hybrid ordinary-parallel speculative decoding strategy guided by a threshold derived from throughput analysis, speculative priority scheduling to preserve draft--target overlap under multi-tenant traffic, and draft-side prompt compression to reduce draft latency. We implement SPECTRE in \texttt{SGLang} and evaluate it across multiple draft--target model pairs, reasoning benchmarks, real-world long-context workloads, and a wide range of batch sizes. Results show that SPECTRE consistently improves large-model serving throughput while causing only minor interference to the native workloads of tail-model services. In large-model deployments, including Qwen3-235B-A22B with TP=8, SPECTRE achieves up to \textbf{2.28$\times$ speedup} over autoregressive decoding and up to an additional \textbf{66\% relative improvement} over the strongest speculative decoding baselines. Talk is cheap, we show you the code: \url{https://github.com/sgl-project/sglang/pull/22272}.
}

%% file: tex/01_intro.tex
\section{Introduction}
\label{sec:intro}

Large language model (LLM) serving platforms\cite{Agrawal2023SARATHIEL,10.1109/ISCA59077.2024.00019} are increasingly deployed as multi-model cloud systems, where shared infrastructure supports models with different sizes, capabilities, and service roles \cite{280768}. In practice, user demand in such systems is often long-tailed: a small number of popular large models receive most requests, while many smaller models in the tail see much lighter traffic \cite{aegaeon}. As these tail models remain online to serve the full model portfolio, their own traffic often falls short of fully utilizing their generation capacity. This imbalance motivates the reuse of idle tail-model capacity to assist heavily loaded large-model services.

In this paper, we investigate this opportunity through speculative decoding, a lossless acceleration method. Crucially, this setting also creates an opportunity to parallelize draft generation and target-side verification, allowing the two stages to overlap \cite{shen2026doublebreakingaccelerationlimit,liu2025pearl} instead of being serialized. 
However, such parallelism is not always beneficial, as it may reduce the acceptance length in speculative decoding, thereby diminishing the effective speedup and making its coordination a central systems challenge.

\begin{figure}
    \centering
    \includegraphics[width=1\linewidth]{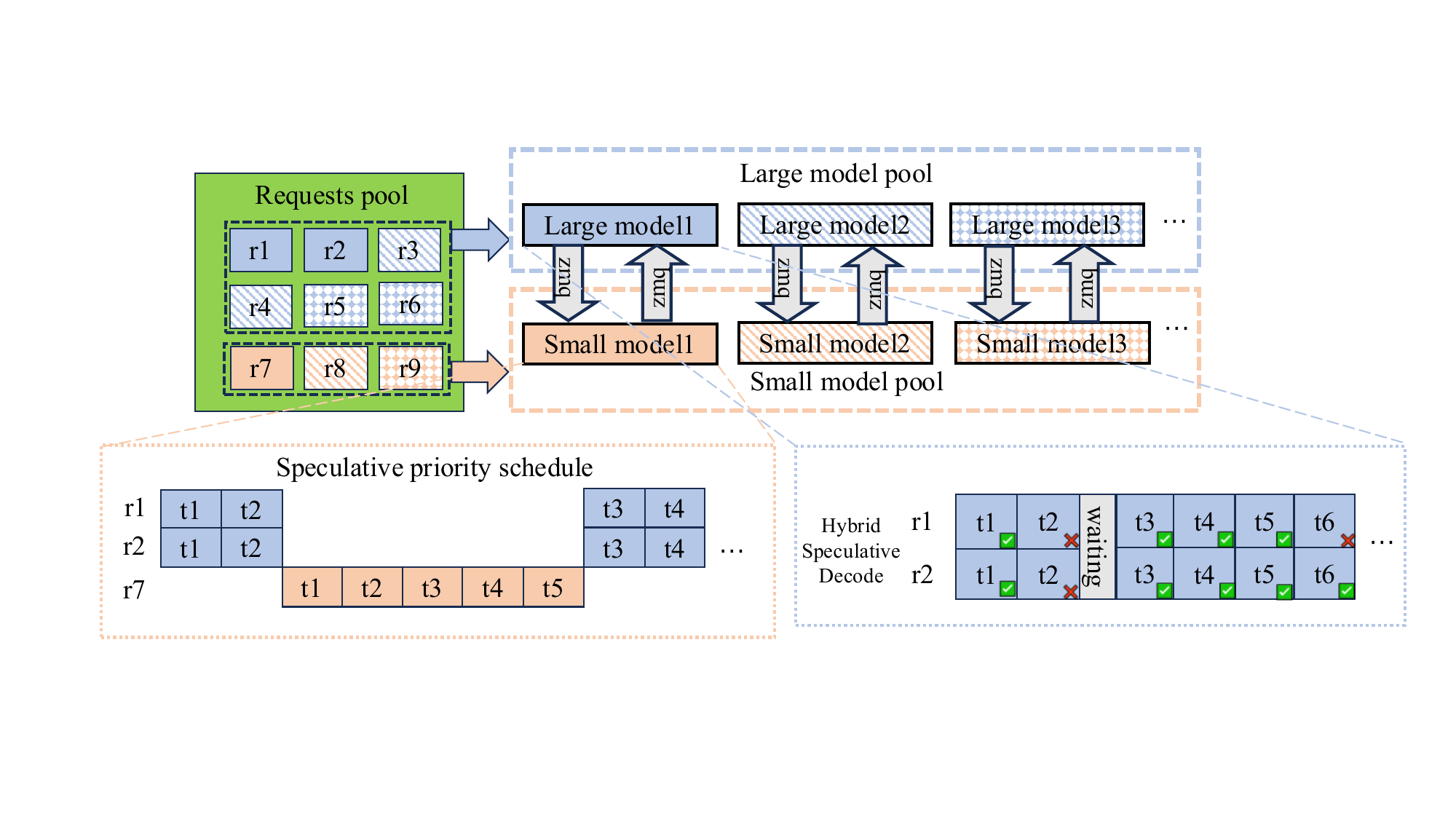}
    \caption{
        A large-model service obtains speculative drafts from an underutilized tail-model service via ZMQ. Blue requests correspond to large-model (target) queries, while orange requests correspond to small-model (draft-side normal traffic) queries. Each request is routed to the service instance with the matching pattern (hatched style). The draft-side system continues serving normal traffic while handling target-issued draft requests under speculative priority scheduling. SPECTRE further applies hybrid ordinary-parallel speculative decoding, dynamically switching modes based on a threshold derived from throughput analysis.
    }
    \label{fig:overview}
\vspace{-5mm}
\end{figure}

To address these challenges, we present \textbf{SPECTRE} (Parallel \textbf{SPEC}ulative Decoding with a Multi-\textbf{T}enant \textbf{RE}mote Drafter), a framework shown in Figure \ref{fig:overview} that reuses underutilized tail-model services as remote drafters for heavily loaded large-model services while allowing these services to continue serving their native workloads. SPECTRE makes such parallelism effective through three techniques. First, it adopts a hybrid ordinary-parallel speculative decoding strategy that switches between the two coordination modes based on a threshold derived from our throughput analysis, using parallel execution only when it remains beneficial for the current batch (\S~\ref{sec:analysis}). Second, because speculative drafts are produced by a shared draft-side system, SPECTRE prioritizes speculative draft requests to preserve effective overlap between drafting and verification under multi-tenant traffic. Third, when target-side verification outpaces draft generation, SPECTRE compresses the draft-side prompt to reduce draft latency (\S~\ref{sec:draft_side_optimization}). Together, these techniques make such parallelism effective in long-tail multi-model serving.

Extensive experiments across multiple draft--target model pairs, reasoning benchmarks, real-world long-context workloads, and a wide range of batch sizes show that SPECTRE consistently improves throughput. We further find that moderate draft-side load causes little degradation to target throughput, supporting the practicality of reusing shared tail-model services as remote drafters. For large-model deployments, including Qwen3-235B-A22B with TP=8, SPECTRE achieves up to \textbf{2.28$\times$ speedup} over autoregressive decoding and up to an additional \textbf{66\% relative improvement} over the strongest speculative decoding baselines. At the system level, these throughput gains translate into up to \textbf{81\% higher revenue} than autoregressive decoding.

In summary, this paper makes the following contributions:
\begin{itemize}
    \item We propose \textbf{SPECTRE}, a serving framework that reuses underutilized tail-model services as remote drafters for heavily loaded large-model services while allowing these services to continue serving their native workloads.
    \item We design a hybrid ordinary-parallel speculative decoding strategy for SPECTRE, whose switching rule is guided by a threshold derived from our throughput analysis so that parallel coordination is used only when it remains beneficial.
   \item We develop a speculative priority scheduling policy and a draft-side prompt compression method that preserve effective draft--target overlap and reduce draft latency, enabling tail-model services to support speculative drafting while continuing to serve their native workloads.
    \item We show through experiments across multiple draft--target model pairs, reasoning benchmarks, real-world long-context workloads, and a wide range of batch sizes that SPECTRE consistently improves large-model serving throughput while causing only minor interference to the native workloads of tail-model services.
\end{itemize}

%% file: tex/03_analysis.tex
\begin{figure}
    \centering
    \includegraphics[width=\linewidth]{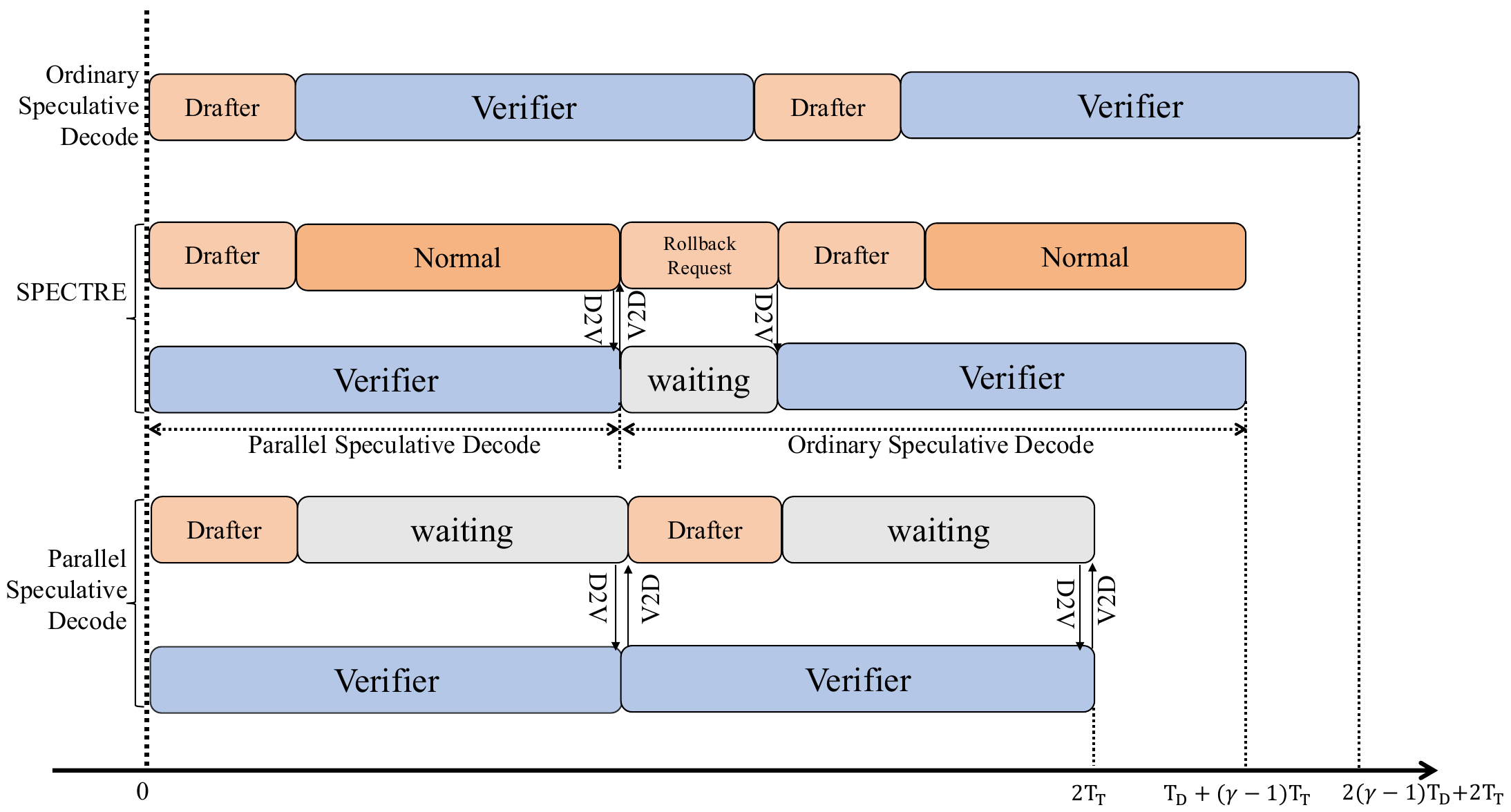}
    \caption{
    Timeline comparison of ordinary speculative decoding, parallel speculative decoding, and SPECTRE. 
    Ordinary speculative decoding is fully serialized. 
    Parallel speculative decoding overlaps draft generation and target verification. 
    SPECTRE adaptively switches between parallel and ordinary coordination.
    }
    \label{fig:pipeline1}
\end{figure}

\section{Throughput-Guided Hybrid Coordination}
\label{sec:analysis}

Figure~\ref{fig:pipeline1} illustrates the difference between ordinary and parallel speculative decoding. In ordinary speculative decoding, the target verification result provides the bonus token that seeds the next draft segment, so each round requires one target verification and only \(\gamma-1\) new draft steps:
\begin{equation}
\mathrm{Thr}_{\mathrm{ord}}
=
\frac{B L}{T_T + (\gamma-1)T_D},
\label{eq:thr_ordinary}
\end{equation}
where \(B\) is the batch size, \(L\) is the accepted length, \(\gamma\) is the number of speculative tokens per verification, and \(T_D\), \(T_T\) denote the latency of one draft step and one target verification. Parallel speculative decoding instead prepares the next draft segment while the current segment is being verified. When \(\gamma T_D<T_T\), draft generation can be hidden by target verification; however, if the current verification rejects the prefix needed by the pre-generated segment, the request falls back to autoregressive decoding. Let \(r\) be this fallback ratio. The parallel throughput is approximated as
\begin{equation}
\mathrm{Thr}_{\mathrm{par}}
\approx
\frac{B\bigl[r+(1-r)L\bigr]}{T_T}.
\label{eq:thr_parallel}
\end{equation}
Comparing the two throughputs gives a critical threshold
\begin{equation}
r^*
=
\frac{(\gamma-1)L T_D}
{\bigl(T_T+(\gamma-1)T_D\bigr)(L-1)}.
\label{eq:r_threshold}
\end{equation}
Parallel coordination is beneficial only when \(r\le r^*\); otherwise, fallback losses outweigh overlap benefits. \textsc{SPECTRE} therefore uses this threshold to switch between parallel and ordinary coordination at runtime, as shown in Fig.~\ref{fig:pipeline1}. The full derivation is provided in Appendix~\ref{app:throughput_analysis}.

%% file: tex/04_method.tex
\begin{figure}[th]
    \centering
    \includegraphics[width=1.0\linewidth]{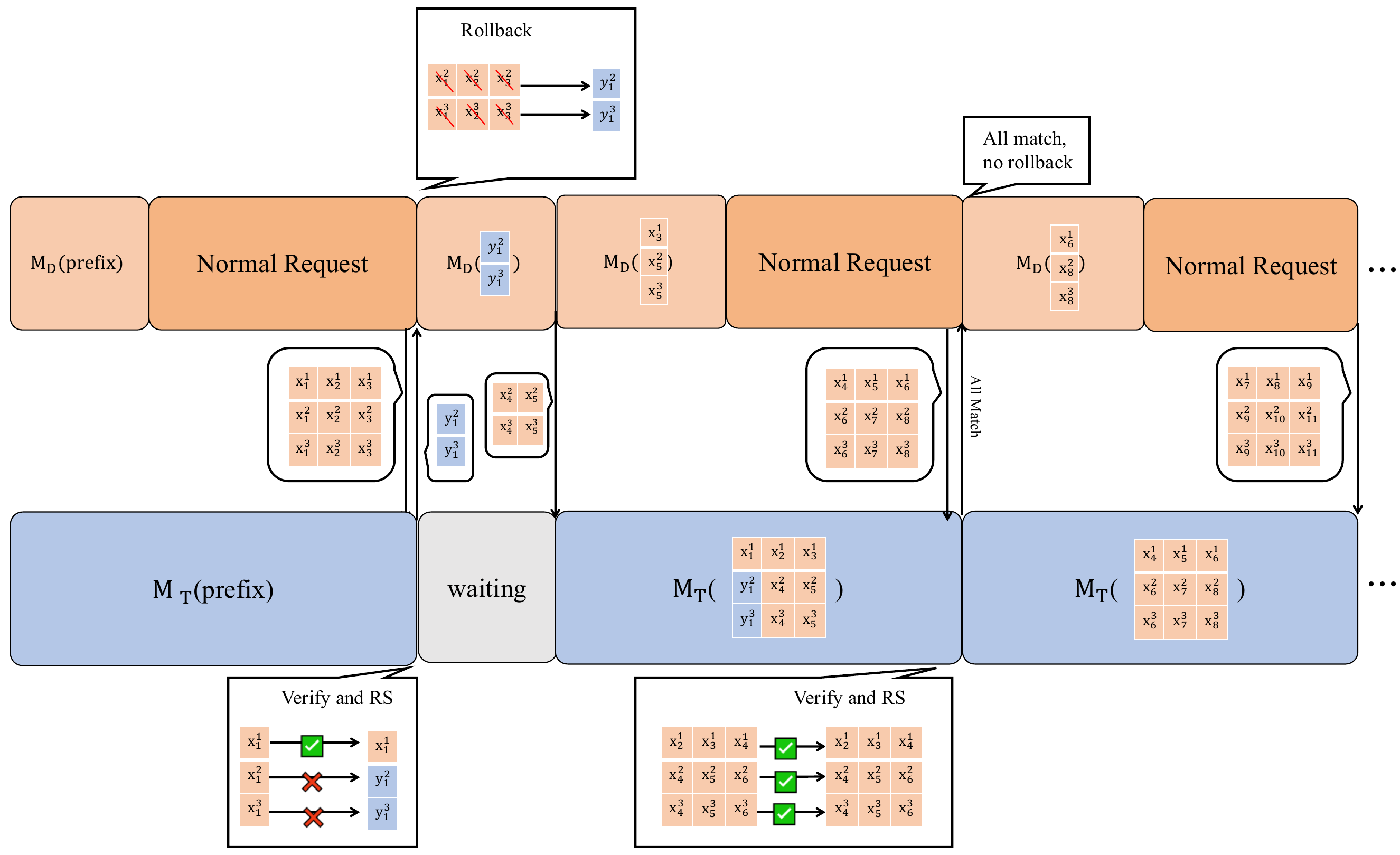}
    \caption{
   Illustration of SPECTRE decoding. 
    The draft model (\(M_D\)) generates candidate tokens while serving normal requests, and the target model (\(M_T\)) verifies them via rejection sampling (RS). 
    When the proportion of rollback requests becomes high, the system switches to ordinary coordination and the target waits for repaired draft tokens; otherwise, decoding proceeds without interruption with overlapped execution. 
    }
    \label{fig:pipeline2}
\end{figure}

\section{Method}

We present \textbf{SPECTRE}, a hybrid ordinary-parallel speculative serving framework for remote draft-target deployment as shown in Figure \ref{fig:pipeline2}. SPECTRE consists of a \emph{target server} hosting the large model and a \emph{draft server} hosting the lightweight draft model. At a high level, SPECTRE combines an \emph{target-side adaptive hybrid policy} with a \emph{draft-side speculative priority scheduler} to coordinate speculative execution under remote serving.

\subsection{Target-Side Adaptive Hybrid Policy}

At each decoding round, the target server selects the execution mode for the next round of the current target-side batch. The decision is made at the batch level and determines how the batch coordinates speculative verification and rollback handling in the next step.

\paragraph{Mode semantics.}

When a request's previous speculative continuation can no longer be directly reused, it enters \emph{rollback}, and the draft server must regenerate an updated continuation. SPECTRE supports two execution modes that differ in whether this draft-side rollback blocks target progress. In \textsc{ordinary} mode, the target launches the next verification pass only after the draft server completes this regeneration. In \textsc{Parallel} mode, the target continues its current computation while the draft server regenerates the updated continuation in parallel.

\paragraph{Rollback-Based Mode Selection.}

After the verification pass at round \(n\), the target performs rejection sampling and identifies the rollback set \(\mathcal{R}_n\). A request enters \(\mathcal{R}_n\) if either its speculative prefix is rejected during target-side verification or its previously prepared draft continuation is invalidated by the verification result. The observed rollback ratio is then
\begin{equation}
\hat r_n = \frac{|\mathcal{R}_n|}{B}.
\end{equation}

The target therefore determines the execution mode for round \(n+1\) according to
\begin{equation}
\mathrm{mode}_{n+1} =
\begin{cases}
\textsc{Parallel}, & \hat r_n \le r^*,\\
\textsc{ordinary}, & \hat r_n > r^*.
\end{cases}
\label{eq:mode_switch_batch}
\end{equation}
where $r^*$ is the threshold derived from \S~\ref{sec:analysis}.

\paragraph{Execution in ordinary mode.}

When $\mathrm{mode}_n=\textsc{ordinary}$, the verification batch of round $n$ is formed only after the rollback requests have been refreshed at the draft side. Let $\mathcal{R}_n$ and $\mathcal{S}_n$ denote the rollback and non-rollback requests at round $n$, respectively.

For each rollback request $i \in \mathcal{R}_n$, the target first sends the latest bonus token $b_n^{(i)}$ to the draft server. Conditioned on this updated prefix, the draft server performs $\gamma-1$ auto-regressive decoding steps and returns a repaired continuation
\begin{equation}
\hat{\mathbf{d}}_n^{(i)}
=
\left[
\hat d_{n,1}^{(i)}, \hat d_{n,2}^{(i)}, \ldots, \hat d_{n,\gamma-1}^{(i)}
\right].
\label{eq:ordinary_repair_seq}
\end{equation}
The target-side candidate sequence used for verification is then
\begin{equation}
\mathbf{y}_n^{(i)}
=
\left[
b_n^{(i)}, \hat{\mathbf{d}}_n^{(i)}
\right],
\qquad i \in \mathcal{R}_n.
\label{eq:ordinary_rollback_seq}
\end{equation}

For each non-rollback request $i \in \mathcal{S}_n$, no refresh is needed, and the target directly reuses the speculative continuation prepared in the previous round,
\begin{equation}
\bar{\mathbf{d}}_n^{(i)}
=
\left[
\bar d_{n,1}^{(i)}, \bar d_{n,2}^{(i)}, \ldots, \bar d_{n,\gamma}^{(i)}
\right].
\label{eq:ordinary_cached_seq}
\end{equation}
Thus,
\begin{equation}
\mathbf{y}_n^{(i)}=\bar{\mathbf{d}}_n^{(i)},
\qquad i \in \mathcal{S}_n.
\label{eq:ordinary_stable_seq}
\end{equation}

The full verification batch is assembled as
\begin{equation}
\mathbf{Y}_n
=
\mathrm{Merge}\!\left(
\{\mathbf{y}_n^{(i)} : i \in \mathcal{R}_n\},
\{\mathbf{y}_n^{(i)} : i \in \mathcal{S}_n\}
\right),
\label{eq:ordinary_merge}
\end{equation}
on which the target performs the verification pass for round $n$. Only after this refreshed batch has been formed and dispatched does the draft server resume speculative preparation for the next round.

\paragraph{Execution in parallel mode.}

When $\mathrm{mode}_n=\textsc{parallel}$, target-side verification and draft-side speculative generation proceed concurrently, without waiting for rollback refresh to complete.

For each non-rollback request $i \in \mathcal{S}_n$, the target reuses the cached speculative continuation from the previous round:
\begin{equation}
\mathbf{y}_n^{(i)}
=
\bar{\mathbf{d}}_n^{(i)}
=
\left[
\bar d_{n,1}^{(i)}, \bar d_{n,2}^{(i)}, \ldots, \bar d_{n,\gamma}^{(i)}
\right],
\qquad i \in \mathcal{S}_n.
\label{eq:parallel_stable_seq}
\end{equation}

For each rollback request $i \in \mathcal{R}_n$, refreshed draft tokens are not yet available. The target therefore constructs a padded candidate sequence using the latest bonus token:
\begin{equation}
\mathbf{p}_n^{(i)}
=
\left[
b_n^{(i)}, \underbrace{\varnothing, \varnothing, \ldots, \varnothing}_{\gamma-1}
\right],
\qquad i \in \mathcal{R}_n,
\label{eq:parallel_rollback_pad}
\end{equation}
where $\varnothing$ denotes padding token with no draft prediction. The target-side candidate is
\begin{equation}
\mathbf{y}_n^{(i)}=\mathbf{p}_n^{(i)},
\qquad i \in \mathcal{R}_n.
\label{eq:parallel_rollback_seq}
\end{equation}

The mixed verification batch is then formed as
\begin{equation}
\mathbf{Y}_n
=
\mathrm{Merge}\!\left(
\{\mathbf{y}_n^{(i)} : i \in \mathcal{R}_n\},
\{\mathbf{y}_n^{(i)} : i \in \mathcal{S}_n\}
\right),
\label{eq:parallel_merge}
\end{equation}
and the target immediately performs the verification pass on $\mathbf{Y}_n$.

Meanwhile, the draft server prepares speculative continuations for the \emph{next} round. To avoid ambiguity with the target-side candidates of the current round, we denote these next-round draft outputs by $\bar{\mathbf{d}}_{n+1}^{(i)}$. Specifically,
\begin{equation}
\bar{\mathbf{d}}_{n+1}^{(i)}
=
\begin{cases}
\left[
\bar d_{n+1,1}^{(i)}, \ldots, \bar d_{n+1,\gamma}^{(i)}
\right], & i \in \mathcal{R}_n, \\[2mm]
\left[
\bar {d'}_{n+1,1}^{(i)}, \ldots, \bar {d'}_{n+1,\gamma}^{(i)}
\right], & i \in \mathcal{S}_n,
\end{cases}
\label{eq:parallel_draft_update}
\end{equation}
where for $i \in \mathcal{R}_n$ generation is conditioned on the updated prefix $b_n^{(i)}$, while for $i \in \mathcal{S}_n$ generation continues from the previously cached speculative state. These continuations $\{\bar{\mathbf{d}}_{n+1}^{(i)}\}$ become available for verification in round $n+1$.

\subsection{Draft-Side Optimization}
\label{sec:draft_side_optimization}
\paragraph{speculative priority scheduling.}
Under remote deployment, the draft server serves both speculative and regular user requests, leading to increased latency under mixed batching. Let $T_D^{\mathrm{mix}}$ denote the per-step draft latency. If
\begin{equation}
\gamma T_D^{\mathrm{mix}} > T_T,
\label{eqn:tle}
\end{equation}
the draft cannot produce $\gamma$ tokens within one verification step, causing the target to stall.
SPECTRE adopts a \emph{speculative priority, non-preemptive} policy that prioritizes speculative requests in the next scheduling round while allowing the current round to complete. 
To prevent normal requests starvation, we enforce a lightweight fairness rule that schedules regular requests after every $K$ speculative steps.

\paragraph{Draft-side context compression.}
When the target is accelerated by advanced inference techniques such as tensor parallelism (TP) \cite{shoeybi2020megatronlmtrainingmultibillionparameter}, draft-side speculative generation may become the runtime bottleneck, weakening draft--target overlap. 
SPECTRE addresses this case by using a streamingLLM-style\cite{streamingLLM} compression on draft-side context during prefill.
Specifically, given an input sequence \(x_{1:S}\) where $S$ is sequence length, the draft retains only a prefix and a suffix,
\begin{equation}
\mathcal{E}(x_{1:S})
=
\left[
x_{:\lfloor \frac{p}{2}S \rfloor},
\; x_{-\lfloor \frac{p}{2}S \rfloor+1:}
\right],
\label{eq:token_eviction}
\end{equation}
where \(p\) is the retained ratio. This compression reduces the effective context length seen by the draft and thereby lowers speculative-generation latency. The resulting context mismatch between draft and target may reduce the accepted length, but can still improve overall target-side throughput when draft latency dominates the critical path.

Despite these optimizations, fully satisfying Eq.~\eqref{eqn:tle} remains challenging in practical serving environments due to system-level variability and workload dynamics. As a result, transient violations of this condition may still occur in certain cases. When such violations happen, the system falls back to a conservative execution mode in which the target waits for the completion of all speculative token generation before verification.

%% file: tex/05_experiment.tex
\section{Experiments}
\label{sec:experiments}
\subsection{Experimental Setup}

\paragraph{Models and deployment.}

We evaluate SPECTRE on three draft--target model pairs: Qwen3-0.6B (TP1) $\rightarrow$ Qwen3-32B (TP1), Qwen3-0.6B (TP1) $\rightarrow$ Qwen3-235B-A22B (TP8) \cite{yang2025qwen3technicalreport}, and DeepSeek-R1-Distill-Qwen-1.5B (TP1) $\rightarrow$ DeepSeek-R1-Distill-Qwen-32B (TP1) \cite{deepseekai2025deepseekv3technicalreport}. The draft and target models are deployed on separate H200 GPUs\cite{nvidia_h200_gpu} and communicate remotely through our serving framework. 
Additional implementation details are provided in Appendix~\ref{app:implementation_details}.

\paragraph{Datasets.}

We evaluate SPECTRE on six datasets: GSM8K \cite{cobbe2021trainingverifierssolvemath}, MATH500 \cite{lightman2024lets}, Minerva Math, LongBench \cite{bai-etal-2024-longbench,bai-etal-2025-longbench}, MRCR \cite{vodrahalli2024michelangelolongcontextevaluations}, and ShareGPT. These datasets cover grade-school math, advanced mathematical reasoning, STEM-oriented problem solving, long-context question answering and retrieval, and conversational data.

\paragraph{Baselines.}
We compare SPECTRE against autoregressive decoding (AR), Standalone \cite{10.5555/3618408.3619203}, EAGLE-3 \cite{li2025eagle}, PEARL \cite{liu2025pearl}, and MineDraft \cite{tang2026minedraftframeworkbatchparallel}, under the same target model, decoding configuration, batch size, and hardware budget. For the Qwen3 series, we use publicly available EAGLE-3 checkpoints\footnote{\url{https://huggingface.co/AngelSlim/Qwen3-32B_eagle3} and \url{https://huggingface.co/lmsys/Qwen3-235B-A22B-EAGLE3}}. We do not report EAGLE-3 results for the DeepSeek series because no public checkpoints are available. For MineDraft, we use the official open-source implementation.\footnote{The official MineDraft implementation is based on \texttt{vLLM 0.9.2}\cite{vllm} and performs relatively poorly in our setup, sometimes even underperforming autoregressive decoding in \texttt{sglang v0.5.7}\cite{sglang}. We attribute this to implementation limitations rather than to the underlying method itself. In addition, MineDraft frequently encounters out-of-memory (OOM) issues under long-context settings, so we exclude it from long-context experiments.} Unless otherwise specified, we use greedy decoding (temperature $=0$), and the target verifies four speculative tokens per round. Detailed settings for each experiment are provided in Appendix~\ref{app:Experiment Settings}.

\input{tables/main_result_compress}
\subsection{TP1 Target Throughput Speedup}

Table~\ref{tab:main_compact_results} reports target throughput speedups for TP1 targets on reasoning datasets (GSM8K, MATH500, and Minerva Math) and long-context datasets (ShareGPT and LongBench) at batch size $=32, 64, 128$. SPECTRE achieves the highest throughput in all reported settings, with up to \textbf{2.28$\times$ speedup} over autoregressive decoding. Compared with prior speculative decoding baselines, SPECTRE delivers more consistent gains across batch sizes and datasets. \textsc{EAGLE3} and \textsc{PEARL} often lose speedup as batch size increases and can even underperform autoregressive decoding on long-context datasets \textsc{MineDraft} is less stable and is excluded from long-context experiments because of out-of-memory issues. In contrast, SPECTRE remains effective across the full set of reported TP1 results. 
The $\Delta$ rows further show that SPECTRE consistently outperforms the SOTA method in each setting, with up to an additional \textbf{66\%} gain on LongBench. Overall, these results show that SPECTRE consistently improves target throughput across both reasoning and long-context datasets. Moreover, we illustrate SPECTRE at small batch size in Appendix.\ref{app:small batch}.

\begin{figure}
    \centering
    \includegraphics[width=0.75\linewidth]{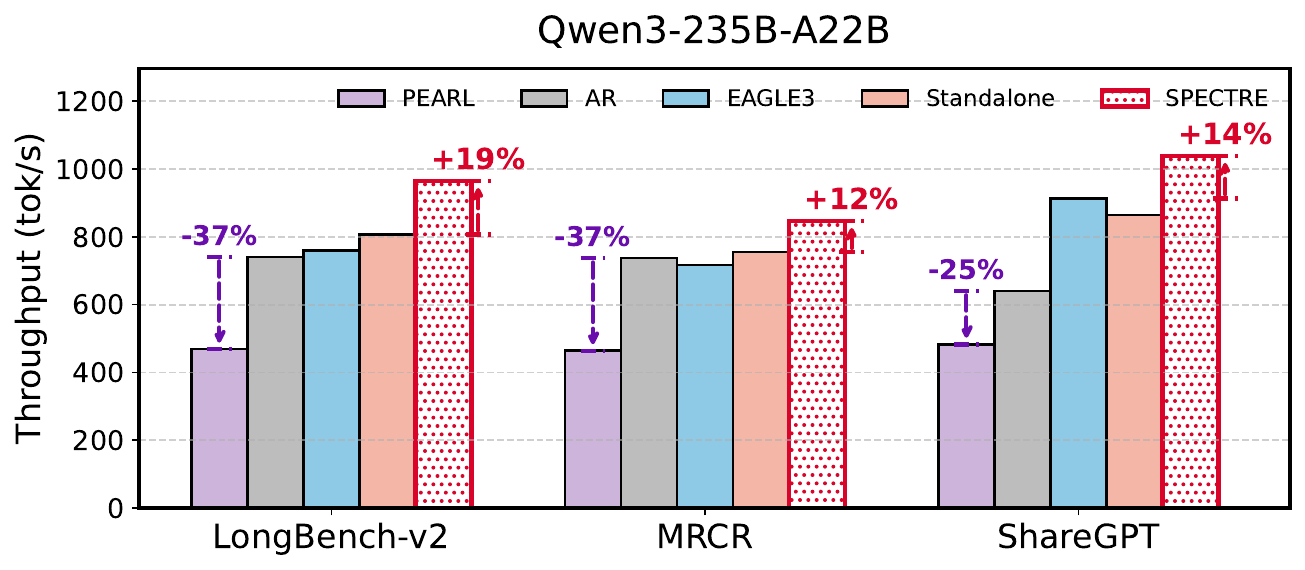}
    \caption{
    Throughput comparison under high-concurrency settings (batch size=128) using Qwen3-235B-A22B (TP=8) as the target model and Qwen3-0.6B (TP=1) as the draft model. We evaluate across three workloads: LongBench-v2, MRCR, and ShareGPT. SPECTRE consistently achieves the highest throughput, outperforming all baselines. The red upward arrows indicate the relative improvement of SPECTRE over the strongest baseline (Standalone for LongBench-v2 and MRCR, and EAGLE3 for ShareGPT), while the purple downward arrows indicate the relative degradation of PEARL compared to the autoregressive baseline (AR). Detailed experimental settings are provided in \ref{app:Experimental setup for Qwen3-235B-A22B}.
}
    \label{fig:qwen3-235b}
\vspace{-3mm}
\end{figure}

\subsection{TP8 Target Throughput Speedup}

Under TP=8 target deployment, target-side verification becomes much faster relative to draft generation, making parallel speculative decoding more difficult to sustain. 
We therefore enable context compression and set $p=0.1$ in \eqref{eq:token_eviction}. 
Figure~\ref{fig:qwen3-235b} reports throughput at batch size 128 on LongBench-v2, MRCR, and ShareGPT. 
\textsc{SPECTRE} consistently achieves the best throughput across all three datasets, exceeding the strongest competing method by \textbf{19\%}, \textbf{12\%}, and \textbf{14\%}, respectively. 
In contrast, existing parallel speculative decoding methods struggle in this regime: \textsc{EAGLE3} and \textsc{Standalone} provide only limited gains, while \textsc{PEARL} even underperforms the autoregressive baseline. 
PEARL degradation stems from the asymmetric compute allocation in our setting, where the target model is deployed with TP=8 while the draft model uses only a single GPU, causing draft generation to become the system bottleneck and preventing effective overlap with target-side verification. 
Overall, these results highlight that naive parallelization is insufficient under high-concurrency and imbalanced resource settings, whereas \textsc{SPECTRE} remains robust and continues to deliver substantial throughput improvements.

\begin{figure}[th]
    \centering
    \includegraphics[width=1\linewidth]{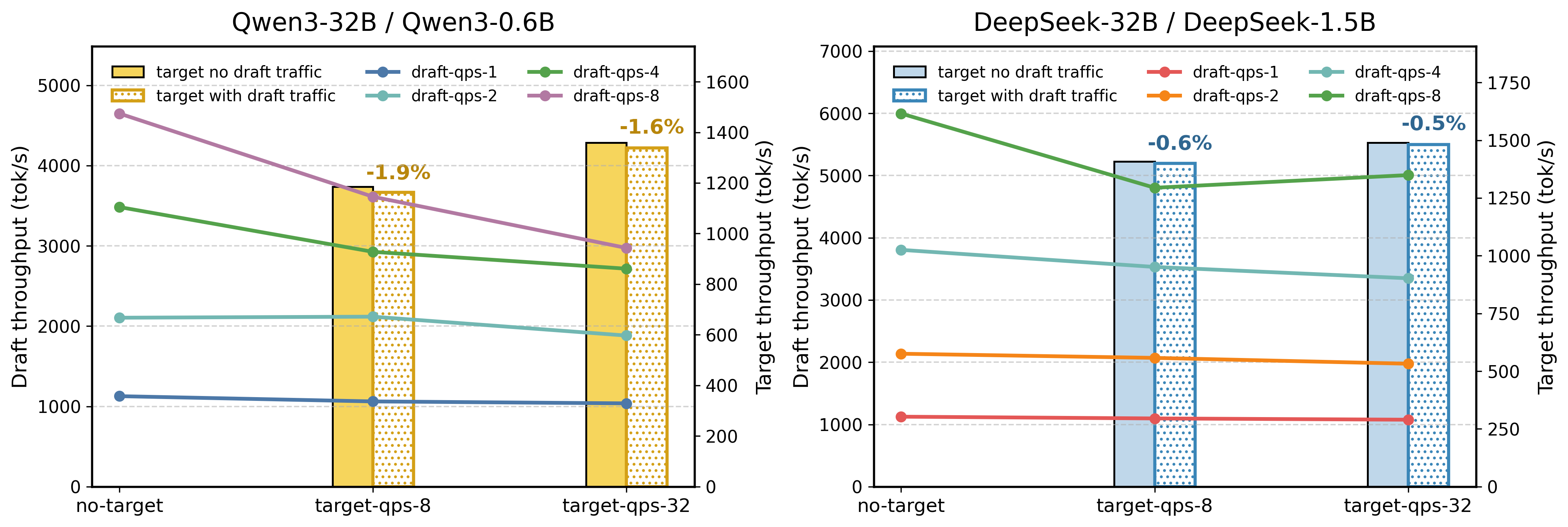}
    \caption{Throughput under mixed draft-side traffic.
The draft model serves both speculative requests from the target model and background user requests (query per second (QPS), denoted by draft-qps-$k$).
The x-axis shows different target request rates.
Solid bars denote target throughput without background draft traffic, while hatched bars denote target throughput with background traffic.
Lines represent draft throughput under different background loads.
The annotated percentages indicate the relative decrease in target throughput when background traffic is present. Detailed experimental settings are provided in \ref{app:mixed_traffic}.
}
    \label{fig:draft traffic}
\vspace{-3mm}
\end{figure}

\begin{table}[th]
\centering
\resizebox{\textwidth}{!}{
\begin{tabular}{lccccc}
\toprule
\textbf{Model Pair} & \textbf{AR} & \textbf{EAGLE3} & \textbf{PEARL} & \textbf{Standalone} & \textbf{SPECTRE} \\
\midrule
Qwen3-32B / Qwen3-0.6B 
& 2.33 & 2.52 & 1.60 & 2.62 & \textbf{2.71 (+1.16$\times$)} \\

Qwen3-235B-A22B / Qwen3-0.6B 
& 0.38 & 0.47 & 0.16 & 0.46 & \textbf{0.48 (+1.26$\times$)} \\

DeepSeek-R1-Distill-Qwen-32B / DeepSeek-R1-Distill-Qwen-1.5B 
& 0.37 & -- & 0.42 & 0.44 & \textbf{0.67 (+1.81$\times$)} \\
\bottomrule
\end{tabular}
}
\caption{
System-level benefit efficiency measured in dollar/1000s/GPU (higher is better).
For SPECTRE, the values in parentheses denote the relative improvement over AR.
SPECTRE consistently achieves the highest benefit efficiency across all model configurations. Price from \url{https://help.aliyun.com/zh/model-studio/model-pricing}. The detail of experiment settings is shown in \ref{app:benefit_setup}.
}
\label{tab:cost_efficiency}
\vspace{-3mm}
\end{table}

\subsection{Performance under shared draft traffic}
\label{sec:shared_draft_traffic}
We next evaluate SPECTRE in a multi-tenant setting where the draft server simultaneously serves background requests and speculative decoding traffic. As shown in Fig.~\ref{fig:draft traffic}, moderate draft-side load has little impact on target throughput. Draft throughput remains nearly unchanged up to 8 QPS and degrades only under heavier load, while target throughput decreases by less than $2\%$ even at draft-QPS 8 and target-QPS 32 for both Qwen3-32B and DeepSeek-R1-Distill-Qwen-32B.
Table~\ref{tab:cost_efficiency} further shows that SPECTRE still delivers the highest system-level benefit in this shared-serving setting, improving from 0.38 to 0.48 ($1.26\times$) on Qwen3-235B-A22B and from 0.37 to 0.67 ($1.81\times$) on DeepSeek-R1-Distill-Qwen-32B. Although draft throughput drops slightly under high load, the overall benefit continues to increase, indicating that SPECTRE can effectively translate additional draft-side utilization into higher end-to-end efficiency. In contrast, \textsc{PEARL} achieves limited gains in this setting, as its draft model occupies a dedicated GPU but does not serve regular user traffic, leading to lower overall resource utilization and reduced system-level benefit.

\input{tables/acc_len}
\vspace{-3mm}
\subsection{Accepted Length Analysis}

Table~\ref{tab:accept_len_avg} reports mean accepted length averaged over batch sizes. We compare SPECTRE with \textsc{EAGLE3}, \textsc{PEARL}, and \textsc{Standalone}. We exclude \textsc{MineDraft} because it changes batch-level scheduling but not token acceptance, and therefore has the same accepted length as \textsc{Standalone}. Across all datasets and models, \textsc{Standalone} achieves the largest accepted length, as expected from ordinary speculative decoding without parallel interference. SPECTRE yields accepted lengths that remain close to \textsc{Standalone}, for example, $2.59$--$3.11$ versus $3.13$--$3.22$ on Qwen3-32B, and $3.00$--$3.61$ versus $3.41$--$3.77$ on DeepSeek-R1-Distill-Qwen-32B. By contrast, \textsc{EAGLE3} and \textsc{PEARL} show noticeably shorter accepted lengths across settings. Together with Table~\ref{tab:main_compact_results}, these results show that SPECTRE attains higher throughput than prior parallel baselines while maintaining accepted lengths closer to \textsc{Standalone}.
\vspace{-3mm}

%% file: tables/main_result_compress.tex
\definecolor{spectrerow}{RGB}{255,245,204}
\definecolor{deltarow}{RGB}{221,235,247}

\begin{table*}[t]
\centering
\scriptsize
\setlength{\tabcolsep}{4pt}
\resizebox{\textwidth}{!}{
\begin{tabular}{llcccccccccc}
\toprule
\textbf{Model (bs)} & \textbf{Method}
& \multicolumn{2}{c}{\textbf{GSM8K}}
& \multicolumn{2}{c}{\textbf{Math500}}
& \multicolumn{2}{c}{\textbf{Minerva Math}}
& \multicolumn{2}{c}{\textbf{ShareGPT}}
& \multicolumn{2}{c}{\textbf{LongBench}} \\
\cmidrule(lr){3-4} \cmidrule(lr){5-6} \cmidrule(lr){7-8} \cmidrule(lr){9-10} \cmidrule(lr){11-12}
& & \textbf{Tok/s} & \textbf{Speedup}
  & \textbf{Tok/s} & \textbf{Speedup}
  & \textbf{Tok/s} & \textbf{Speedup}
  & \textbf{Tok/s} & \textbf{Speedup}
  & \textbf{Tok/s} & \textbf{Speedup} \\
\midrule

\multirow{7}{*}{\makecell{\textbf{Qwen3-32B} \\ \textbf{(bs=32)}}}
& AR         & 1009.48 & 1.00$\times$ & 1001.40 & 1.00$\times$ & 1003.11 & 1.00$\times$ & 753.44 & 1.00$\times$ & 308.06 & 1.00$\times$ \\
& EAGLE3     & 1380.38 & 1.37$\times$ & 1436.39 & 1.43$\times$ & 1283.62 & 1.28$\times$ & 878.34 & 1.17$\times$ & 248.30 & 0.81$\times$ \\
& PEARL      & 1255.99 & 1.24$\times$ & 1020.93 & 1.02$\times$ & 1241.92 & 1.23$\times$ & 945.24 & 1.25$\times$ & \underline{532.38} & \underline{1.73$\times$} \\
& MineDraft  & 759.50  & 0.75$\times$ & 832.34  & 0.83$\times$ & 782.61  & 0.78$\times$ & OOM     & OOM            & OOM     & OOM            \\
& Standalone & \underline{1559.03} & \underline{1.54$\times$} & \underline{1594.78} & \underline{1.59$\times$} & \underline{1588.71} & \underline{1.58$\times$} & \underline{993.60} & \underline{1.32$\times$} & 488.57 & 1.59$\times$ \\
\rowcolor{spectrerow}
& \textbf{SPECTRE} & \textbf{1764.78} & \textbf{1.75$\times$} & \textbf{1728.68} & \textbf{1.73$\times$} & \textbf{1776.95} & \textbf{1.77$\times$} & \textbf{1185.06} & \textbf{1.57$\times$} & \textbf{653.69} & \textbf{2.12$\times$} \\
\rowcolor{deltarow}
& $\Delta$ (\%) &  & $\uparrow 13.6\%$ &  & $\uparrow 8.8\%$ &  & $\uparrow 12.0\%$ &  & $\uparrow 18.9\%$ &  & $\uparrow 22.5\%$ \\
\midrule

\multirow{7}{*}{\makecell{\textbf{Qwen3-32B} \\ \textbf{(bs=64)}}}
& AR         & 1432.60 & 1.00$\times$ & 1423.07 & 1.00$\times$ & 1422.39 & 1.00$\times$ & 806.95 & 1.00$\times$ & 310.59 & 1.00$\times$ \\
& EAGLE3     & 1782.63 & 1.24$\times$ & 1797.62 & 1.26$\times$ & 1776.18 & 1.25$\times$ & 959.11 & 1.19$\times$ & 235.85 & 0.76$\times$ \\
& PEARL      & 1483.45 & 1.04$\times$ & 754.85  & 0.53$\times$ & 828.45  & 0.58$\times$ & \underline{1038.49} & \underline{1.29$\times$} & 495.32 & 1.59$\times$ \\
& MineDraft  & 1193.47 & 0.83$\times$ & 1295.63 & 0.91$\times$ & 1231.83 & 0.87$\times$ & OOM     & OOM            & OOM     & OOM            \\
& Standalone & \underline{1923.28} & \underline{1.34$\times$} & \underline{1945.95} & \underline{1.37$\times$} & \underline{1924.57} & \underline{1.35$\times$} & 1015.15 & 1.26$\times$ & \underline{517.73} & \underline{1.67$\times$} \\
\rowcolor{spectrerow}
& \textbf{SPECTRE} & \textbf{2183.14} & \textbf{1.52$\times$} & \textbf{2148.39} & \textbf{1.51$\times$} & \textbf{2152.47} & \textbf{1.51$\times$} & \textbf{1353.11} & \textbf{1.68$\times$} & \textbf{706.81} & \textbf{2.28$\times$} \\
\rowcolor{deltarow}
& $\Delta$ (\%) &  & $\uparrow 13.4\%$ &  & $\uparrow 10.2\%$ &  & $\uparrow 11.9\%$ &  & $\uparrow 30.2\%$ &  & $\uparrow 36.5\%$ \\
\midrule

\multirow{7}{*}{\makecell{\textbf{Qwen3-32B} \\ \textbf{(bs=128)}}}
& AR         & 1762.27 & 1.00$\times$ & 1762.52 & 1.00$\times$ & 1590.46 & 1.00$\times$ & 789.93 & 1.00$\times$ & 465.29 & 1.00$\times$ \\
& EAGLE3     & 1930.57 & 1.10$\times$ & 1941.57 & 1.10$\times$ & 1774.24 & 1.12$\times$ & 982.93 & 1.24$\times$ & 302.95 & 0.65$\times$ \\
& PEARL      & 893.30  & 0.51$\times$ & 637.58  & 0.36$\times$ & 842.30  & 0.53$\times$ & \underline{1069.28} & \underline{1.35$\times$} & 483.91 & 1.04$\times$ \\
& MineDraft  & 1556.27 & 0.88$\times$ & 1544.90 & 0.88$\times$ & 1409.60 & 0.89$\times$ & OOM     & OOM            & OOM     & OOM            \\
& Standalone & \underline{2037.20} & \underline{1.16$\times$} & \underline{2053.80} & \underline{1.17$\times$} & \underline{1951.21} & \underline{1.23$\times$} & 1048.48 & 1.33$\times$ & \underline{515.68} & \underline{1.11$\times$} \\
\rowcolor{spectrerow}
& \textbf{SPECTRE} & \textbf{2430.48} & \textbf{1.38$\times$} & \textbf{2421.28} & \textbf{1.37$\times$} & \textbf{2348.38} & \textbf{1.48$\times$} & \textbf{1359.09} & \textbf{1.72$\times$} & \textbf{800.45} & \textbf{1.72$\times$} \\
\rowcolor{deltarow}
& $\Delta$ (\%) &  & $\uparrow 19.0\%$ &  & $\uparrow 17.1\%$ &  & $\uparrow 20.3\%$ &  & $\uparrow 27.4\%$ &  & $\uparrow 55.0\%$ \\
\midrule

\multirow{6}{*}{\makecell{\textbf{DS-32B} \\ \textbf{(bs=32)}}}
& AR         & 1025.18 & 1.00$\times$ & 1017.38 & 1.00$\times$ & 1014.42 & 1.00$\times$ & 800.39 & 1.00$\times$ & 449.11 & 1.00$\times$ \\
& PEARL      & 1407.74 & 1.37$\times$ & 1403.23 & 1.38$\times$ & 1429.11 & 1.41$\times$ & 456.09 & 0.57$\times$ & 202.83 & 0.45$\times$ \\
& MineDraft  & 957.13  & 0.93$\times$ & 1003.12 & 0.99$\times$ & 992.49  & 0.98$\times$ & OOM     & OOM            & OOM     & OOM            \\
& Standalone & \underline{2137.28} & \underline{2.08$\times$} & \underline{2147.52} & \underline{2.11$\times$} & \underline{2131.65} & \underline{2.10$\times$} & \underline{925.76} & \underline{1.16$\times$} & \underline{565.95} & \underline{1.26$\times$} \\
\rowcolor{spectrerow}
& \textbf{SPECTRE} & \textbf{2268.75} & \textbf{2.21$\times$} & \textbf{2259.68} & \textbf{2.22$\times$} & \textbf{2271.10} & \textbf{2.24$\times$} & \textbf{1408.17} & \textbf{1.76$\times$} & \textbf{833.92} & \textbf{1.86$\times$} \\
\rowcolor{deltarow}
& $\Delta$ (\%) &  & $\uparrow 6.2\%$ &  & $\uparrow 5.2\%$ &  & $\uparrow 6.7\%$ &  & $\uparrow 52.1\%$ &  & $\uparrow 47.6\%$ \\
\midrule

\multirow{6}{*}{\makecell{\textbf{DS-32B} \\ \textbf{(bs=64)}}}
& AR         & 1431.09 & 1.00$\times$ & 1428.67 & 1.00$\times$ & 1412.98 & 1.00$\times$ & 841.87 & 1.00$\times$ & 478.90 & 1.00$\times$ \\
& PEARL      & 1681.16 & 1.17$\times$ & 1050.97 & 0.74$\times$ & 988.95  & 0.70$\times$ & 449.22 & 0.53$\times$ & 197.65 & 0.41$\times$ \\
& MineDraft  & 1462.13 & 1.02$\times$ & 1475.78 & 1.03$\times$ & 1465.81 & 1.04$\times$ & OOM     & OOM            & OOM     & OOM            \\
& Standalone & \underline{2741.32} & \underline{1.92$\times$} & \underline{2147.52} & \underline{1.50$\times$} & \underline{2753.48} & \underline{1.95$\times$} & \underline{954.79} & \underline{1.13$\times$} & \underline{592.64} & \underline{1.24$\times$} \\
\rowcolor{spectrerow}
& \textbf{SPECTRE} & \textbf{2806.23} & \textbf{1.96$\times$} & \textbf{2884.79} & \textbf{2.02$\times$} & \textbf{2867.57} & \textbf{2.03$\times$} & \textbf{1274.28} & \textbf{1.50$\times$} & \textbf{887.01} & \textbf{1.85$\times$} \\
\rowcolor{deltarow}
& $\Delta$ (\%) &  & $\uparrow 2.1\%$ &  & $\uparrow 34.7\%$ &  & $\uparrow 4.1\%$ &  & $\uparrow 32.7\%$ &  & $\uparrow 49.2\%$ \\
\midrule

\multirow{6}{*}{\makecell{\textbf{DS-32B} \\ \textbf{(bs=128)}}}
& AR         & 1766.62 & 1.00$\times$ & 1758.49 & 1.00$\times$ & 1584.39 & 1.00$\times$ & 826.19 & 1.00$\times$ & 480.42 & 1.00$\times$ \\
& PEARL      & 911.46  & 0.52$\times$ & 637.67  & 0.36$\times$ & 950.51  & 0.60$\times$ & 937.30 & 1.13$\times$ & 218.80 & 0.46$\times$ \\
& MineDraft  & 1625.18 & 0.92$\times$ & 1621.92 & 0.92$\times$ & 1552.44 & 0.98$\times$ & OOM     & OOM            & OOM     & OOM            \\
& Standalone & \underline{2885.66} & \underline{1.63$\times$} & \underline{2906.47} & \underline{1.65$\times$} & \underline{2753.25} & \underline{1.74$\times$} & \underline{988.23} & \underline{1.20$\times$} & \underline{599.22} & \underline{1.25$\times$} \\
\rowcolor{spectrerow}
& \textbf{SPECTRE} & \textbf{3131.47} & \textbf{1.77$\times$} & \textbf{3218.42} & \textbf{1.83$\times$} & \textbf{3029.84} & \textbf{1.91$\times$} & \textbf{1489.56} & \textbf{1.80$\times$} & \textbf{998.38} & \textbf{2.08$\times$} \\
\rowcolor{deltarow}
& $\Delta$ (\%) &  & $\uparrow 8.6\%$ &  & $\uparrow 10.9\%$ &  & $\uparrow 9.8\%$ &  & $\uparrow 50.0\%$ &  & $\uparrow 66.4\%$ \\
\bottomrule
\end{tabular}
}
\caption{
Throughput (Tok/s) and speedup over AR for each model, batch size, and benchmark. For each benchmark, The best results are marked in bold and the SOTA results are underlined for each benchmark.
Light yellow rows correspond to SPECTRE, and light blue rows show SPECTRE's relative speedup gain over the SOTA method.
Missing MineDraft entries indicate out-of-memory (OOM) errors.
The detail of experiment settings is shown in \ref{app:Detailed_Experimental_Setup_for_Table}.
}
\label{tab:main_compact_results}
\end{table*}

%% file: tables/acc_len.tex
\begin{table}[t]
\centering
\begin{tabular}{lccccc}
\toprule
Dataset & AR & EAGLE3 & PEARL & Standalone & SPECTRE \\
\midrule

\multicolumn{6}{c}{\textbf{Qwen3-32B}} \\
\midrule
GSM8K & 1.00 & 1.96 & 1.96 & \textbf{3.17} & 2.61 \\
Math500 & 1.00 & 1.96 & 1.66 & \textbf{3.18} & 2.61 \\
Minerva Math & 1.00 & 1.90 & 1.75 & \textbf{3.18} & 2.59 \\
ShareGPT & 1.00 & 2.44 & 2.03 & \textbf{3.13} & 3.11 \\
Longbench & 1.00 & 1.90 & 2.05 & \textbf{3.22} & 2.68 \\

\midrule
\multicolumn{6}{c}{\textbf{DeepSeek-R1-Distill-Qwen-32B}} \\
\midrule
GSM8K & 1.00 & -- & 2.19 & \textbf{3.41} & 3.00 \\
Math500 & 1.00 & -- & 2.01 & \textbf{3.47} & 3.07 \\
Minerva Math & 1.00 & -- & 1.99 & \textbf{3.47} & 3.02 \\
ShareGPT & 1.00 & -- & 2.16 & \textbf{3.57} & 3.27 \\
Longbench & 1.00 & -- & 1.91 & \textbf{3.77} & 3.61 \\

\bottomrule
\end{tabular}
\caption{
Mean accepted length averaged over batch sizes $\{32,64,128\}$. 
Detailed experimental settings are provided in  \ref{app:Detailed_Experimental_Setup_for_Table}.
}
\label{tab:accept_len_avg}
\vspace{-5mm}
\end{table}

%% file: tex/06_conclusion.tex
\section{Conclusion}
\label{sec:conclusion}
\vspace{-3mm}
We present \textbf{SPECTRE}, a hybrid ordinary-parallel speculative decoding framework that improves the efficiency of LLM inference by addressing system-level inefficiencies in existing approaches. Experiments across multiple benchmarks show that SPECTRE consistently achieves the highest throughput, delivering up to $2.28\times$ speedup over \textsc{AR} and up to $66\%$ improvement over \textsc{Standalone}, while maintaining acceptance lengths close to the latter. These results suggest that parallel speculative decode alone is insufficient for high performance, and that jointly considering acceptance behavior and execution efficiency is critical. Overall, SPECTRE provides an effective direction for scalable and efficient LLM inference.

%% file: tex/A0_implementation_details.tex
\appendix

\section{Implementation Details}
\label{app:implementation_details}

\paragraph{System overview.}
SPECTRE separates speculative decoding into two cooperating services: a \emph{target} server and a \emph{draft} server. The target server is responsible for maintaining the authoritative generation state and for performing verification with the large model. The draft server runs asynchronously and predicts future tokens that may be accepted by the target. The two servers exchange compact synchronization messages over a lightweight messaging layer. Each message is tagged with both a request identifier and a decoding-round identifier so that delayed or duplicated messages can be safely ignored.

\subsection{Target side}

The target server maintains the authoritative decoding state for every request and is the only component allowed to commit output target tokens. For request $i$ at round $n$, let $\mathbf{p}_n^{(i)}$ denote the committed prefix, and let $\mathbf{d}_n^{(i)}$ denote the speculative tokens currently available at the target. The target then forms a verification candidate
\begin{equation}
\mathbf{y}_n^{(i)} = \mathrm{Assemble}\!\left(\mathbf{p}_n^{(i)}, \mathbf{d}_n^{(i)}\right),
\end{equation}
and submits the batch
\begin{equation}
\mathbf{Y}_n = \{\mathbf{y}_n^{(i)}\}_{i=1}^B
\end{equation}
to the target model for verification.

Meanwhile, the target sends speculative-generation requests to the remote draft server. Let
\begin{equation}
\mathbf{q}_n^{(i)} = (i,n,\mathbf{p}_n^{(i)})
\end{equation}
denote the draft query issued for request $i$ at round $n$, where $(i,n)$ is the request-round identifier. The draft server returns a speculative tokens
\begin{equation}
\tilde{\mathbf{d}}_n^{(i)} = [\tilde d_{n,1}^{(i)},\ldots,\tilde d_{n,\gamma}^{(i)}].
\end{equation}
Because target verification and draft generation proceed concurrently, the target-side forward pass overlaps with both network communication and draft-side computation.

After verification completes, the target computes the accepted length $L_n^{(i)}$ and commits the verified tokens
\begin{equation}
\mathbf{p}_{n+1}^{(i)} = \mathrm{Commit}\!\left(\mathbf{p}_n^{(i)}, \mathbf{y}_n^{(i)}, L_n^{(i)}\right).
\end{equation}
The target then determines whether the draft tokens remains reusable. If the verified tokens is still consistent with the speculative suffix, the target retains the unmatched suffix for the next round; otherwise, it discards the stale tokens. Formally, letting $\mathrm{Suffix}(\mathbf{d},\ell)$ denote the suffix of $\mathbf{d}$ after the first $\ell$ accepted tokens, the next-round cached tokens is
\begin{equation}
\mathbf{d}_{n+1}^{(i)}
=
\begin{cases}
\mathrm{Suffix}\!\left(\mathbf{d}_n^{(i)}, L_n^{(i)}\right), & \text{if } \mathrm{Consistent}\!\left(\mathbf{p}_{n+1}^{(i)}, \mathbf{d}_n^{(i)}\right)=1, \\[1mm]
\emptyset, & \text{otherwise}.
\end{cases}
\label{eq:target_suffix_reuse}
\end{equation}
Thus, draft tokens are treated purely as proposals; only target-verified tokens can enter the committed prefix.

\paragraph{Unified verification interface.}
The target does not introduce a separate verification algorithm for remote drafts. 
Instead, it converts remote speculative tokens into the tree-structured verification input used by \texttt{sglang} speculative decoding. 
Let $\mathcal{T}_n^{(i)}$ denote the verification tree associated with request $i$ at round $n$. In the common case of a single remote tokens, $\mathcal{T}_n^{(i)}$ degenerates into a chain of length $\gamma$ and verified by the same target-side procedure. Hence, SPECTRE reuses the existing verification machinery and modifies only the source of speculative proposals.

\paragraph{Asynchronous reply handling.}
Remote draft replies may be delayed, reordered, or correspond to stale decoding states. Therefore, each reply carries the identifier $(i,n)$. A returned tokens $\tilde{\mathbf{d}}_n^{(i)}$ is accepted only if both its request identity and round identity match the current target-side state:
\begin{equation}
\mathrm{Valid}\!\left(\tilde{\mathbf{d}}_n^{(i)}\right)=1
\iff
(i,n)=(i_n^\star,n_n^\star),
\label{eq:reply_validity}
\end{equation}
where $(i_n^\star,n_n^\star)$ is the active request-round pair maintained by the target. Replies failing Eq.~\eqref{eq:reply_validity} are dropped. This ensures correctness under asynchronous arrivals and fluctuating network delay.

\paragraph{Fallback and circuit breaker.}
To prevent remote instability from reducing serving throughput, the target includes two fallback mechanisms. First, if the target-side scheduler determines that speculative expansion should be disabled, the request falls back to standard one-token decoding, i.e.,
\begin{equation}
\mathbf{d}_n^{(i)}=\emptyset
\quad\Longrightarrow\quad
\mathbf{y}_n^{(i)}=\mathbf{p}_n^{(i)} \oplus [\text{one target-decoded token}],
\end{equation}
where $\oplus$ denotes concatenation. Second, the target maintains a circuit breaker for the remote draft service. Let $m_n$ be an indicator of whether the draft reply for round $n$ arrives before the timeout threshold, and let
\begin{equation}
c_n =
\begin{cases}
c_{n-1}+1, & m_n=0,\\
c_{n-1}, & m_n=1.
\end{cases}
\end{equation}
If $c_n \ge C_{\max}$, the target disables remote speculation for a cooldown period of $H$ rounds:
\begin{equation}
\mathbf{d}_{n+h}^{(i)}=\emptyset,\qquad h=1,\ldots,H.
\end{equation}
After the cooldown period, the target probes the remote draft service again and re-enables speculation only if timely replies are restored. This design keeps the target-side execution stable even when the remote draft service is temporarily unavailable or communication becomes unreliable.

\subsection{Draft side}

The draft server is implemented as a persistent speculative state machine. 
For each active request $i$, it maintains a local state
\begin{equation}
\mathcal{S}_n^{(i)} = \left(\mathbf{h}_n^{(i)}, \mathbf{d}_n^{(i)}, \mathbf{k}_n^{(i)}\right),
\end{equation}
where $\mathbf{h}_n^{(i)}$ is the current decoded tokens, $\mathbf{d}_n^{(i)}$ is the speculative tokens, and $\mathbf{k}_n^{(i)}$ denotes the KV cache. 
This state persists across multiple synchronization rounds.

\paragraph{Message handling and update.}
At round $n$, the draft receives a synchronization message
\begin{equation}
\mathbf{m}_n^{(i)} = (i,n,\mathbf{p}_n^{(i)}),
\end{equation}
where $\mathbf{p}_n^{(i)}$ is the verified prefix from the target. 
If multiple messages for the same request accumulate, only the latest one is retained.

The draft reconciles the target prefix $\mathbf{p}_n^{(i)}$ with its local decoded tokens $\mathbf{h}_n^{(i)}$ by finding the first divergence point
\begin{equation}
\delta_n^{(i)} = \min \{ t : \mathbf{p}_n^{(i)}[t] \neq \mathbf{h}_n^{(i)}[t] \}.
\end{equation}

\paragraph{State recovery.}
Based on $\delta_n^{(i)}$, the draft updates its state as follows:
\begin{equation}
\mathcal{S}_n^{(i)} \leftarrow
\begin{cases}
\mathcal{S}_n^{(i)}, & \delta_n^{(i)} = \gamma \quad (\text{no divergence}), \\[1mm]
\mathrm{Rollback}\!\left(\mathcal{S}_n^{(i)}, \delta_n^{(i)}\right), & \text{otherwise} 
\end{cases}
\end{equation}
where $\mathrm{Rollback}$ truncates the speculative suffix and frees the corresponding KV cache, while $\mathrm{Rebuild}$ reconstructs the state from the verified prefix.

\paragraph{Speculative generation.}
After reconciliation, the draft generates a speculative tokens of length $\gamma$:
\begin{equation}
\mathbf{d}_{n+1}^{(i)} = \mathrm{Decode}_D\!\left(\mathbf{p}_n^{(i)}, \mathbf{k}_n^{(i)}, \gamma\right),
\end{equation}
and returns it to the target. The request is then paused with its KV cache preserved:
\begin{equation}
\mathcal{S}_{n+1}^{(i)} \leftarrow \mathrm{Pause}\!\left(\mathcal{S}_n^{(i)}\right).
\end{equation}

\paragraph{Prompt compression.}
When the target-side compute significantly exceeds the draft-side capacity such that $\gamma T_D \gtrsim T_T$, the draft applies prompt compression to reduce latency. 
Given an input prompt of length $S$, the compressed prompt is
\begin{equation}
\mathbf{p}' = \left[\mathbf{p}_{:\lfloor \frac{p}{2}S \rfloor},\; \mathbf{p}_{-\lfloor \frac{p}{2}S \rfloor+1:}\right].
\end{equation}
The draft then performs decoding conditioned on $\mathbf{p}'$ instead of the full prompt, reducing prefill and decoding cost.

\paragraph{Fair scheduling.}
To prevent starvation of regular user requests, the draft enforces a lightweight fairness constraint. 
Let $c_n$ denote the number of consecutive speculative decoding steps. The scheduler updates
\begin{equation}
cnt_n =
\begin{cases}
cnt_{n-1}+1, & \text{speculative step},\\
0, & \text{regular step}.
\end{cases}
\end{equation}
When $cnt_n \ge K$, the scheduler prioritizes one round of regular requests:
\begin{equation}
\text{if } cnt_n \ge K \;\Rightarrow\; \text{schedule regular requests and reset } cnt_n \leftarrow 0,
\end{equation}
with $K=10$.

\paragraph{Latency mismatch.}
Due to the variability of real-world systems, the condition
\begin{equation}
\gamma T_D^{\mathrm{mix}} > T_T
\end{equation}
may still occur. In this case, the system switches to a conservative mode where the target waits for the draft to finish generating the full speculative sequence before verification. 
Since draft generation becomes the bottleneck, the per-round latency is dominated by the draft side and can be approximated as
\begin{equation}
T_{\mathrm{round}} \approx \gamma T_D^{\mathrm{mix}}.
\end{equation}
This ensures correct execution while avoiding unstable behavior under severe latency mismatch.

\subsection{Communication layer}

SPECTRE uses a custom asynchronous messaging backend built on ZeroMQ. The target acts as a central router, while each draft server acts as a dealer endpoint with its own identity. This routing model allows the target to direct requests to a specific draft worker and to track which draft workers are currently alive.

To support both single-node and multi-node deployments efficiently, the communication layer uses local inter-process transport on a single machine and TCP transport across machines. The implementation also separates normal data traffic from control traffic. In particular, draft workers periodically send heartbeat messages to the target, and the target removes workers from its active registry if their heartbeat expires. This provides a simple liveness mechanism without requiring an external coordination service.

The messaging backend is implemented as a multi-threaded C++ extension rather than a pure Python queue. The goal is to minimize Python overhead on the critical path. Sending, receiving, unpacking, and endpoint monitoring are performed by separate background threads, while Python interacts only with already unpacked request batches. Messages are serialized in batches using a compact binary format, which substantially reduces framing overhead compared with per-request transmission.

The transport layer also includes explicit defenses against pathological runtime conditions. It uses bounded queues to avoid unbounded memory growth under backpressure, discards stale buffered messages that have exceeded a timeout, and records malformed payloads and shutdown-time drops for debugging. These details are not algorithmically central, but they are important for making the system reliable in long-running serving workloads where network delay, process restarts, and temporary overload are unavoidable.

Overall, SPECTRE is implemented as a pipelined remote speculative system in which the target server owns correctness and verification, the draft server maintains persistent speculative state with efficient rollback, and the communication layer provides asynchronous, version-aware, and fault-tolerant transport. The practical efficiency of the system comes from overlapping communication with target-side computation, reusing paused draft states across rounds, and restricting expensive cache rebuilds to the cases where the target and draft histories truly diverge.

\input{tex/02_related_work}

%% file: tex/02_related_work.tex
\section{Related Work}

\subsection{Speculative Decoding}

Speculative decoding accelerates auto-regressive generation by letting a lightweight draft model propose multiple candidate tokens and using a larger target model to verify them with fewer target-side decoding steps. Canonical methods preserve the target distribution through rejection sampling, and subsequent work improves draft quality, verification efficiency, and proposal policies within this sequential draft-then-verify framework~\citep{10.5555/3618408.3619203,chen2023accelerating,xia-etal-2023-speculative,yan-etal-2025-decoding,hong2026inferencecostaware,zhang2026racerretrievalaugmentedcontextualrapid,eagle, eagle2, miao2023specinfer,medusa}.

More recent work increases concurrency between drafting and verification instead of treating them as strictly sequential stages. 
PEARL reduces the bubbles between successive verification steps through pre-verify and post-verify scheduling, thereby alleviating the mutual waiting between drafting and verification \cite{liu2025pearl}. SSD pushes this further by predicting likely verification outcomes and pre-speculating for multiple such outcomes during an ongoing verification \cite{kumar2026speculative}. MineDraft studies batch-parallel speculative execution to hide drafting latency under verification and improve overlap at the system level~\citep{tang2026minedraftframeworkbatchparallel}. Our method is most closely related to recent work on parallel speculative decoding, and introduces a hybrid ordinary-parallel design for batched target-side execution.

\subsection{LLM Serving and Resource Utilization}

A growing body of systems work studies LLM serving through better scheduling, disaggregation, and resource orchestration to improve GPU utilization under heterogeneous and time-varying workloads~\citep{aegaeon,10.5555/3691938.3691949,10.5555/3691938.3691948,windserve,10.1145/3779212.3790135,10.5555/3692070.3692543,10.1145/3773772,10.5555/3691938.3691947}. These works are motivated by the observation that serving load is often uneven across requests, execution phases, and co-deployed models, leaving some resources underutilized under static provisioning. Among them, Weaver is closest to our motivation: it identifies hot/cold demand skew in multi-LLM serving and exploits idle capacity from cold models to assist hot models through attention offloading~\citep{weaver,yu2025prismunleashinggpusharing}. Our work is similarly motivated by skewed utilization in multi-model serving, but uses lightly loaded small-model services to provide speculative drafts for heavily loaded large-model services.

%% file: tex/A_analysis.tex
\section{Detailed Throughput Analysis}
\label{app:throughput_analysis}

This appendix provides the detailed derivation of the throughput models used in Section~\ref{sec:analysis}. We consider the decode stage with batch size \(B\). Let \(T_D\) and \(T_T\) denote the latency of one draft-model forward step and one target-model verification pass, respectively. Let \(L\) denote the accepted length, and let \(\gamma\) denote the number of speculative tokens processed in one target verification. We focus on the regime
\begin{equation}
\gamma T_D < T_T,
\label{eq:app_overlap_condition}
\end{equation}
where draft-side generation can potentially be overlapped with target-side verification.

\paragraph{Ordinary speculative decoding.}
As shown in the top pipeline of Fig.~\ref{fig:pipeline1}, ordinary speculative decoding executes draft generation and target verification sequentially across rounds. After initialization, each round verifies a speculative segment of length \(\gamma\) and produces one target-side bonus token. Since this bonus token becomes the first token of the next speculative segment, the draft model only needs to generate \(\gamma-1\) new speculative tokens in each subsequent round. Therefore, the per-round latency is
\begin{equation}
T_{\mathrm{ord}}
=
T_T + (\gamma - 1)T_D.
\end{equation}
Since each round produces \(B L\) accepted tokens for a batch of size \(B\), the ordinary speculative decoding throughput is
\begin{equation}
\mathrm{Thr}_{\mathrm{ord}}
=
\frac{B L}{T_{\mathrm{ord}}}
=
\frac{B L}{T_T + (\gamma-1)T_D}.
\label{eq:app_thr_ordinary}
\end{equation}

\paragraph{Parallel speculative decoding.}
As shown in the bottom pipeline of Fig.~\ref{fig:pipeline1}, parallel speculative decoding overlaps target verification and draft generation across different rounds. At round \(n\), the target verifies the speculative segment prepared in round \(n-1\), while the draft simultaneously prepares the speculative segment for round \(n+1\). Unlike ordinary speculative decoding, the draft model must generate \(\gamma\) new speculative tokens in each round, rather than \(\gamma-1\), because the first token of the next segment is generated before the current target verification result is available.

Let \(r\) be the fraction of requests whose pre-generated speculative continuation becomes invalid after verification. These requests cannot reuse the prepared continuation and fall back to autoregressive decoding, contributing \(Br\) tokens per batch. The remaining fraction \(1-r\) successfully reuses the prepared speculative segment and contributes \(B(1-r)L\) tokens per batch. Thus, the total number of output tokens per round is
\begin{equation}
B\bigl[r + (1-r)L\bigr].
\end{equation}
Under Eq.~\eqref{eq:app_overlap_condition}, draft-side generation is hidden by target-side verification, so the per-round latency is approximated by
\begin{equation}
T_{\mathrm{par}} \approx T_T.
\end{equation}
The throughput of parallel speculative decoding is therefore
\begin{equation}
\mathrm{Thr}_{\mathrm{par}}
=
\frac{B\bigl[r + (1-r)L\bigr]}{T_{\mathrm{par}}}
\approx
\frac{B\bigl[r + (1-r)L\bigr]}{T_T}.
\label{eq:app_thr_parallel}
\end{equation}

\paragraph{Critical fallback ratio.}
We derive the condition under which ordinary speculative decoding outperforms parallel speculative decoding:
\begin{equation}
\mathrm{Thr}_{\mathrm{ord}} > \mathrm{Thr}_{\mathrm{par}}.
\end{equation}
Substituting Eq.~\eqref{eq:app_thr_ordinary} and Eq.~\eqref{eq:app_thr_parallel}, and cancelling \(B\), gives
\begin{equation}
\frac{L}{T_T + (\gamma-1)T_D}
>
\frac{r + (1-r)L}{T_T}.
\end{equation}
Since \(r + (1-r)L = L - r(L-1)\), we have
\begin{align}
L T_T
&>
\bigl(L - r(L-1)\bigr)
\bigl(T_T + (\gamma-1)T_D\bigr) \\
&=
L T_T + L(\gamma-1)T_D
-
r(L-1)\bigl(T_T + (\gamma-1)T_D\bigr).
\end{align}
Rearranging yields
\begin{equation}
r(L-1)\bigl(T_T + (\gamma-1)T_D\bigr)
>
L(\gamma-1)T_D.
\end{equation}
Assuming \(L>1\), ordinary speculative decoding is preferable when
\begin{equation}
r
>
\frac{(\gamma-1)L T_D}
{\bigl(T_T + (\gamma-1)T_D\bigr)(L-1)}
=
r^*.
\label{eq:app_r_threshold}
\end{equation}
Equivalently, parallel speculative decoding is beneficial only when \(r \le r^*\). This threshold captures the core trade-off in Fig.~\ref{fig:pipeline1}: parallel execution hides draft latency, but its benefit disappears when too many pre-generated continuations become invalid. \textsc{SPECTRE} therefore uses this condition to guide its hybrid ordinary-parallel execution strategy.

%% file: tex/A1_more_experiment.tex
\input{tables/more_res}
\input{tables/temp06}
\section{Additional Experiment}
\label{app:more_experiment}
\subsection{Experiment for small batch size}
\label{app:small batch}
We report results for small batch sizes (bs=1 and 16) in Table~\ref{tab:appendix_small_bs}. 
At bs=1, SPECTRE consistently achieves the best performance across all datasets and both model families.
On Qwen3-32B, speedups ranging from 1.74$\times$ to 3.12$\times$ and outperform the SOTA by up to 30.0\% on ShareGPT at bs=1,
On DeepSeek-R1-Distill-Qwen-32B, speedups ranging  from 1.89$\times$ to 2.97$\times$ and outperform the SOTA by up to 44.8\% on LongBench at bs=16. 
This indicates that even in the latency-sensitive regime where speculative parallelism is limited, SPECTRE can still effectively utilize draft-side computation.
We also observe that some baselines degrade significantly in this regime. For example, EAGLE3 drops below 1.0$\times$ on LongBench at bs=16, and MineDraft consistently underperforms due to system inefficiencies. In contrast, SPECTRE exhibits stable gains across both reasoning and long-context workloads. 
Overall, these results demonstrate that SPECTRE not only scales well to large batch sizes (as shown in table.~\ref{tab:main_compact_results}), but also delivers robust and consistent improvements in small-batch settings.

\subsection{Throughput at Temperature 0.6}
\label{app:temp06_results}

Table~\ref{tab:temp06_all_bs_results} reports throughput results under sampling temperature $0.6$, which we adopt to follow the default generation configurations, ensuring a realistic evaluation setting. 
Overall, \textsc{SPECTRE} consistently achieves strong performance, with particularly notable gains over SOTA on ShareGPT and LongBench (up to $50.9\%$ and $41.9\%$ when target model is DeepSeek-R1-Distill-Qwen-32B, bs=128), demonstrating its outstanding performance. 
As batch size increases, the advantage of \textsc{SPECTRE} becomes more pronounced: for Qwen3-32B, improvements grow from marginal gains at $\text{bs}=1$ (e.g., $+0.5\%$ on GSM8K) to substantial gains at larger batch sizes (e.g., $+20.7\%$ at $\text{bs}=64$), while for DeepSeek-R1-Distill-Qwen-32B, gains exceed $34.1\%$ on LongBench at $\text{bs}\geq32$, indicating superior scalability in high-throughput serving regimes. 
These results collectively validate that \textsc{SPECTRE} maintains consistent throughput improvements and favorable scaling behavior under realistic sampling conditions.

%% file: tables/more_res.tex
\begin{table*}[t]
\centering
\scriptsize
\setlength{\tabcolsep}{4pt}
\resizebox{\textwidth}{!}{
\begin{tabular}{llcccccccccc}
\toprule
\textbf{Model (bs)} & \textbf{Method}
& \multicolumn{2}{c}{\textbf{GSM8K}}
& \multicolumn{2}{c}{\textbf{Math500}}
& \multicolumn{2}{c}{\textbf{Minerva Math}}
& \multicolumn{2}{c}{\textbf{ShareGPT}}
& \multicolumn{2}{c}{\textbf{LongBench}} \\
\cmidrule(lr){3-4} \cmidrule(lr){5-6} \cmidrule(lr){7-8} \cmidrule(lr){9-10} \cmidrule(lr){11-12}
& & \textbf{Tok/s} & \textbf{Speedup}
  & \textbf{Tok/s} & \textbf{Speedup}
  & \textbf{Tok/s} & \textbf{Speedup}
  & \textbf{Tok/s} & \textbf{Speedup}
  & \textbf{Tok/s} & \textbf{Speedup} \\
\midrule

\multirow{7}{*}{\makecell{\textbf{Qwen3-32B} \\ \textbf{(bs=1)}}}
& AR         & 51.20 & 1.00$\times$ & 51.17 & 1.00$\times$ & 51.15 & 1.00$\times$ & 47.12 & 1.00$\times$ & 47.64 & 1.00$\times$ \\
& EAGLE3     & 86.54 & 1.69$\times$ & 84.07 & 1.64$\times$ & 83.26 & 1.63$\times$ & 72.17 & 1.53$\times$ & 75.37 & 1.58$\times$ \\
& PEARL      & 92.13 & 1.80$\times$ & 90.59 & 1.77$\times$ & 96.26 & 1.88$\times$ & 109.67 & 2.33$\times$ & 89.48 & 1.88$\times$ \\
& MineDraft  & 49.53 & 0.97$\times$ & 50.21 & 0.98$\times$ & 49.71 & 0.97$\times$ & -- & -- & -- & -- \\
& Standalone & \underline{117.10} & \underline{2.29$\times$} & \underline{115.03} & \underline{2.25$\times$} & \underline{118.71} & \underline{2.32$\times$} & \underline{112.92} & \underline{2.40$\times$} & \underline{106.65} & \underline{2.24$\times$} \\
\rowcolor{spectrerow}
& \textbf{SPECTRE} & \textbf{120.81} & \textbf{2.36$\times$} & \textbf{120.25} & \textbf{2.35$\times$} & \textbf{121.07} & \textbf{2.37$\times$} & \textbf{146.96} & \textbf{3.12$\times$} & \textbf{109.33} & \textbf{2.29$\times$} \\
\rowcolor{deltarow}
& $\Delta$ (\%) &  & $\uparrow 3.1\%$ &  & $\uparrow 4.4\%$ &  & $\uparrow 2.2\%$ &  & $\uparrow 30.0\%$ &  & $\uparrow 2.2\%$ \\
\midrule

\multirow{7}{*}{\makecell{\textbf{Qwen3-32B} \\ \textbf{(bs=16)}}}
& AR         & 632.18 & 1.00$\times$ & 629.20 & 1.00$\times$ & 627.41 & 1.00$\times$ & 548.85 & 1.00$\times$ & 347.39 & 1.00$\times$ \\
& EAGLE3     & 966.58 & 1.53$\times$ & 968.45 & 1.54$\times$ & 891.35 & 1.42$\times$ & 693.24 & 1.26$\times$ & 280.41 & 0.81$\times$ \\
& PEARL      & 897.05 & 1.42$\times$ & 919.35 & 1.46$\times$ & 859.95 & 1.37$\times$ & 843.96 & 1.54$\times$ & 515.64 & 1.48$\times$ \\
& MineDraft  & 456.31 & 0.72$\times$ & 479.99 & 0.76$\times$ & 450.68 & 0.72$\times$ & -- & -- & -- & -- \\
& Standalone & \underline{1122.98} & \underline{1.78$\times$} & \underline{1094.45} & \underline{1.74$\times$} & \underline{1118.89} & \underline{1.78$\times$} & \underline{933.13} & \underline{1.70$\times$} & \underline{558.48} & \underline{1.61$\times$} \\
\rowcolor{spectrerow}
& \textbf{SPECTRE} & \textbf{1261.15} & \textbf{1.99$\times$} & \textbf{1294.87} & \textbf{2.06$\times$} & \textbf{1172.97} & \textbf{1.87$\times$} & \textbf{1133.64} & \textbf{2.07$\times$} & \textbf{605.73} & \textbf{1.74$\times$} \\
\rowcolor{deltarow}
& $\Delta$ (\%) &  & $\uparrow 11.8\%$ &  & $\uparrow 18.4\%$ &  & $\uparrow 5.1\%$ &  & $\uparrow 21.8\%$ &  & $\uparrow 8.1\%$ \\
\midrule

\multirow{6}{*}{\makecell{\textbf{DS-32B} \\ \textbf{(bs=1)}}}
& AR         & 52.37 & 1.00$\times$ & 52.56 & 1.00$\times$ & 52.55 & 1.00$\times$ & 49.63 & 1.00$\times$ & 47.78 & 1.00$\times$ \\
& PEARL      & 107.03 & 2.04$\times$ & 108.42 & 2.06$\times$ & 108.82 & 2.07$\times$ & 113.76 & 2.29$\times$ & 61.26 & 1.28$\times$ \\
& MineDraft  & 60.97 & 1.16$\times$ & 64.46 & 1.23$\times$ & 64.39 & 1.23$\times$ & -- & -- & -- & -- \\
& Standalone & \underline{119.00} & \underline{2.27$\times$} & \underline{123.99} & \underline{2.36$\times$} & \underline{124.49} & \underline{2.37$\times$} & \underline{121.66} & \underline{2.45$\times$} & \underline{119.64} & \underline{2.50$\times$} \\
\rowcolor{spectrerow}
& \textbf{SPECTRE} & \textbf{139.62} & \textbf{2.67$\times$} & \textbf{142.48} & \textbf{2.71$\times$} & \textbf{141.52} & \textbf{2.69$\times$} & \textbf{147.60} & \textbf{2.97$\times$} & \textbf{131.99} & \textbf{2.76$\times$} \\
\rowcolor{deltarow}
& $\Delta$ (\%) &  & $\uparrow 17.6\%$ &  & $\uparrow 14.8\%$ &  & $\uparrow 13.5\%$ &  & $\uparrow 21.2\%$ &  & $\uparrow 10.4\%$ \\
\midrule

\multirow{6}{*}{\makecell{\textbf{DS-32B} \\ \textbf{(bs=16)}}}
& AR         & 623.74 & 1.00$\times$ & 623.46 & 1.00$\times$ & 621.02 & 1.00$\times$ & 588.30 & 1.00$\times$ & 389.52 & 1.00$\times$ \\
& PEARL      & 992.22 & 1.59$\times$ & 1002.69 & 1.61$\times$ & 930.88 & 1.50$\times$ & 617.22 & 1.05$\times$ & 186.68 & 0.48$\times$ \\
& MineDraft  & 563.63 & 0.90$\times$ & 591.26 & 0.95$\times$ & 581.86 & 0.94$\times$ & -- & -- & -- & -- \\
& Standalone & \underline{1362.28} & \underline{2.18$\times$} & \underline{1420.58} & \underline{2.28$\times$} & \underline{1371.41} & \underline{2.21$\times$} & \underline{888.04} & \underline{1.51$\times$} & \underline{558.48} & \underline{1.43$\times$} \\
\rowcolor{spectrerow}
& \textbf{SPECTRE} & \textbf{1469.75} & \textbf{2.36$\times$} & \textbf{1539.59} & \textbf{2.47$\times$} & \textbf{1383.45} & \textbf{2.23$\times$} & \textbf{1111.64} & \textbf{1.89$\times$} & \textbf{805.51} & \textbf{2.07$\times$} \\
\rowcolor{deltarow}
& $\Delta$ (\%) &  & $\uparrow 8.3\%$ &  & $\uparrow 8.3\%$ &  & $\uparrow 0.9\%$ &  & $\uparrow 25.2\%$ &  & $\uparrow 44.8\%$ \\

\bottomrule
\end{tabular}
}
\caption{
Results for small batch sizes (bs=1 and bs=16).
We report throughput (Tok/s) and speedup over AR. 
SPECTRE consistently achieves the best performance across all settings.
}
\label{tab:appendix_small_bs}
\end{table*}

%% file: tables/temp06.tex
\definecolor{spectrerow}{RGB}{255,245,204}
\definecolor{deltarow}{RGB}{221,235,247}

\begin{table*}[ht]
\centering
\scriptsize
\setlength{\tabcolsep}{3.8pt}
\renewcommand{\arraystretch}{1.05}
\resizebox{\textwidth}{!}{
\begin{tabular}{llcccccccccc}
\toprule
\textbf{Model (bs)} & \textbf{Method}
& \multicolumn{2}{c}{\textbf{GSM8K}}
& \multicolumn{2}{c}{\textbf{Math500}}
& \multicolumn{2}{c}{\textbf{Minerva Math}}
& \multicolumn{2}{c}{\textbf{ShareGPT}}
& \multicolumn{2}{c}{\textbf{LongBench}} \\
\cmidrule(lr){3-4} \cmidrule(lr){5-6} \cmidrule(lr){7-8} \cmidrule(lr){9-10} \cmidrule(lr){11-12}
& & \textbf{Tok/s} & \textbf{AR $\times$}
  & \textbf{Tok/s} & \textbf{AR $\times$}
  & \textbf{Tok/s} & \textbf{AR $\times$}
  & \textbf{Tok/s} & \textbf{AR $\times$}
  & \textbf{Tok/s} & \textbf{AR $\times$} \\
\midrule

\multirow{6}{*}{\makecell{\textbf{Qwen3-32B} \\ \textbf{(bs=1)}}}
& AR         & 51.39 & 1.00$\times$ & 51.58 & 1.00$\times$ & 51.57 & 1.00$\times$ & 50.96 & 1.00$\times$ & 49.12 & 1.00$\times$ \\
& EAGLE3     & 81.47 & 1.59$\times$ & 88.01 & 1.71$\times$ & 79.20 & 1.54$\times$ & 100.99 & 1.98$\times$ & 74.58 & 1.52$\times$ \\
& PEARL      & 84.55 & 1.65$\times$ & 95.36 & 1.85$\times$ & 91.66 & 1.78$\times$ & 85.77 & 1.68$\times$ & 81.65 & 1.66$\times$ \\
& Standalone & \underline{113.29} & \underline{2.20$\times$} & \underline{114.72} & \underline{2.22$\times$} & \underline{106.11} & \underline{2.06$\times$} & \underline{109.61} & \underline{2.15$\times$} & \underline{96.13} & \underline{1.96$\times$} \\
& \cellcolor{spectrerow}\textbf{SPECTRE} & \cellcolor{spectrerow}\textbf{113.80} & \cellcolor{spectrerow}\textbf{2.21$\times$} & \cellcolor{spectrerow}\textbf{114.89} & \cellcolor{spectrerow}\textbf{2.23$\times$} & \cellcolor{spectrerow}\textbf{120.50} & \cellcolor{spectrerow}\textbf{2.34$\times$} & \cellcolor{spectrerow}\textbf{137.75} & \cellcolor{spectrerow}\textbf{2.70$\times$} & \cellcolor{spectrerow}\textbf{96.32} & \cellcolor{spectrerow}\textbf{1.96$\times$} \\
& \cellcolor{deltarow}$\Delta$ & \cellcolor{deltarow} & \cellcolor{deltarow}$\uparrow 0.5\%$ & \cellcolor{deltarow} & \cellcolor{deltarow}$\uparrow 0.4\%$ & \cellcolor{deltarow} & \cellcolor{deltarow}$\uparrow 13.6\%$ & \cellcolor{deltarow} & \cellcolor{deltarow}$\uparrow 25.6\%$ & \cellcolor{deltarow} & \cellcolor{deltarow}$\uparrow 0.0\%$ \\
\midrule

\multirow{6}{*}{\makecell{\textbf{Qwen3-32B} \\ \textbf{(bs=16)}}}
& AR         & 674.52 & 1.00$\times$ & 673.22 & 1.00$\times$ & 673.21 & 1.00$\times$ & 556.52 & 1.00$\times$ & 429.98 & 1.00$\times$ \\
& EAGLE3     & 1007.62 & 1.49$\times$ & 1000.59 & 1.49$\times$ & 901.82 & 1.34$\times$ & \underline{816.34} & \underline{1.47$\times$} & \underline{527.65} & \underline{1.23$\times$} \\
& PEARL      & 808.18 & 1.20$\times$ & 845.22 & 1.26$\times$ & 769.13 & 1.14$\times$ & 607.71 & 1.09$\times$ & 490.99 & 1.14$\times$ \\
& Standalone & \underline{1064.48} & \underline{1.58$\times$} & \underline{1104.52} & \underline{1.64$\times$} & \underline{1058.97} & \underline{1.57$\times$} & 756.03 & 1.36$\times$ & 504.17 & 1.17$\times$ \\
& \cellcolor{spectrerow}\textbf{SPECTRE} & \cellcolor{spectrerow}\textbf{1168.70} & \cellcolor{spectrerow}\textbf{1.73$\times$} & \cellcolor{spectrerow}\textbf{1232.49} & \cellcolor{spectrerow}\textbf{1.83$\times$} & \cellcolor{spectrerow}\textbf{1158.37} & \cellcolor{spectrerow}\textbf{1.72$\times$} & \cellcolor{spectrerow}\textbf{829.70} & \cellcolor{spectrerow}\textbf{1.49$\times$} & \cellcolor{spectrerow}\textbf{624.43} & \cellcolor{spectrerow}\textbf{1.45$\times$} \\
& \cellcolor{deltarow}$\Delta$ & \cellcolor{deltarow} & \cellcolor{deltarow}$\uparrow 9.5\%$ & \cellcolor{deltarow} & \cellcolor{deltarow}$\uparrow 11.6\%$ & \cellcolor{deltarow} & \cellcolor{deltarow}$\uparrow 9.5\%$ & \cellcolor{deltarow} & \cellcolor{deltarow}$\uparrow 1.4\%$ & \cellcolor{deltarow} & \cellcolor{deltarow}$\uparrow 17.8\%$ \\
\midrule

\multirow{6}{*}{\makecell{\textbf{Qwen3-32B} \\ \textbf{(bs=32)}}}
& AR         & 1132.42 & 1.00$\times$ & 1125.59 & 1.00$\times$ & 1121.91 & 1.00$\times$ & 836.60 & 1.00$\times$ & 583.38 & 1.00$\times$ \\
& EAGLE3     & 1456.27 & 1.29$\times$ & 1521.70 & 1.35$\times$ & 1468.74 & 1.31$\times$ & \underline{1064.58} & \underline{1.27$\times$} & \underline{675.56} & \underline{1.16$\times$} \\
& PEARL      & 1169.81 & 1.03$\times$ & 1147.89 & 1.02$\times$ & 1136.12 & 1.01$\times$ & 749.78 & 0.90$\times$ & 521.78 & 0.89$\times$ \\
& Standalone & \underline{1528.09} & \underline{1.35$\times$} & \underline{1562.91} & \underline{1.39$\times$} & \underline{1542.32} & \underline{1.37$\times$} & 752.51 & 0.90$\times$ & 498.76 & 0.85$\times$ \\
& \cellcolor{spectrerow}\textbf{SPECTRE} & \cellcolor{spectrerow}\textbf{1732.45} & \cellcolor{spectrerow}\textbf{1.53$\times$} & \cellcolor{spectrerow}\textbf{1838.71} & \cellcolor{spectrerow}\textbf{1.63$\times$} & \cellcolor{spectrerow}\textbf{1745.68} & \cellcolor{spectrerow}\textbf{1.56$\times$} & \cellcolor{spectrerow}\textbf{1119.59} & \cellcolor{spectrerow}\textbf{1.34$\times$} & \cellcolor{spectrerow}\textbf{718.28} & \cellcolor{spectrerow}\textbf{1.23$\times$} \\
& \cellcolor{deltarow}$\Delta$ & \cellcolor{deltarow} & \cellcolor{deltarow}$\uparrow 13.3\%$ & \cellcolor{deltarow} & \cellcolor{deltarow}$\uparrow 17.3\%$ & \cellcolor{deltarow} & \cellcolor{deltarow}$\uparrow 13.9\%$ & \cellcolor{deltarow} & \cellcolor{deltarow}$\uparrow 5.5\%$ & \cellcolor{deltarow} & \cellcolor{deltarow}$\uparrow 6.0\%$ \\
\midrule

\multirow{6}{*}{\makecell{\textbf{Qwen3-32B} \\ \textbf{(bs=64)}}}
& AR         & 1692.39 & 1.00$\times$ & 1690.82 & 1.00$\times$ & 1679.18 & 1.00$\times$ & 904.59 & 1.00$\times$ & 608.44 & 1.00$\times$ \\
& EAGLE3     & \underline{1962.39} & \underline{1.16$\times$} & 1343.24 & 0.79$\times$ & \underline{1866.97} & \underline{1.11$\times$} & \underline{1212.35} & \underline{1.34$\times$} & \underline{697.02} & \underline{1.15$\times$} \\
& PEARL      & 991.17 & 0.59$\times$ & 1386.43 & 0.82$\times$ & 1341.77 & 0.80$\times$ & 835.76 & 0.92$\times$ & 480.88 & 0.79$\times$ \\
& Standalone & 1846.31 & 1.09$\times$ & \underline{1852.00} & \underline{1.10$\times$} & 1798.87 & 1.07$\times$ & 778.18 & 0.86$\times$ & 529.35 & 0.87$\times$ \\
& \cellcolor{spectrerow}\textbf{SPECTRE} & \cellcolor{spectrerow}\textbf{2372.58} & \cellcolor{spectrerow}\textbf{1.40$\times$} & \cellcolor{spectrerow}\textbf{2336.98} & \cellcolor{spectrerow}\textbf{1.38$\times$} & \cellcolor{spectrerow}\textbf{2206.57} & \cellcolor{spectrerow}\textbf{1.31$\times$} & \cellcolor{spectrerow}\textbf{1237.14} & \cellcolor{spectrerow}\textbf{1.37$\times$} & \cellcolor{spectrerow}\textbf{759.21} & \cellcolor{spectrerow}\textbf{1.25$\times$} \\
& \cellcolor{deltarow}$\Delta$ & \cellcolor{deltarow} & \cellcolor{deltarow}$\uparrow 20.7\%$ & \cellcolor{deltarow} & \cellcolor{deltarow}$\uparrow 25.5\%$ & \cellcolor{deltarow} & \cellcolor{deltarow}$\uparrow 18.0\%$ & \cellcolor{deltarow} & \cellcolor{deltarow}$\uparrow 2.2\%$ & \cellcolor{deltarow} & \cellcolor{deltarow}$\uparrow 8.7\%$ \\
\midrule

\multirow{6}{*}{\makecell{\textbf{Qwen3-32B} \\ \textbf{(bs=128)}}}
& AR         & 2190.45 & 1.00$\times$ & 2194.08 & 1.00$\times$ & 1923.54 & 1.00$\times$ & 878.22 & 1.00$\times$ & 603.63 & 1.00$\times$ \\
& EAGLE3     & \underline{2192.51} & \underline{1.00$\times$} & \underline{2204.53} & \underline{1.00$\times$} & \underline{1993.04} & \underline{1.04$\times$} & \underline{1273.50} & \underline{1.45$\times$} & 528.53 & 0.88$\times$ \\
& PEARL      & 637.69 & 0.29$\times$ & 887.87 & 0.40$\times$ & 817.81 & 0.43$\times$ & 933.93 & 1.06$\times$ & 472.93 & 0.78$\times$ \\
& Standalone & 1930.04 & 0.88$\times$ & 1962.45 & 0.89$\times$ & 1860.39 & 0.97$\times$ & 838.17 & 0.95$\times$ & \underline{702.76} & \underline{1.16$\times$} \\
& \cellcolor{spectrerow}\textbf{SPECTRE} & \cellcolor{spectrerow}\textbf{2608.87} & \cellcolor{spectrerow}\textbf{1.19$\times$} & \cellcolor{spectrerow}\textbf{2659.45} & \cellcolor{spectrerow}\textbf{1.21$\times$} & \cellcolor{spectrerow}\textbf{2426.95} & \cellcolor{spectrerow}\textbf{1.26$\times$} & \cellcolor{spectrerow}\textbf{1313.67} & \cellcolor{spectrerow}\textbf{1.50$\times$} & \cellcolor{spectrerow}\textbf{754.00} & \cellcolor{spectrerow}\textbf{1.25$\times$} \\
& \cellcolor{deltarow}$\Delta$ & \cellcolor{deltarow} & \cellcolor{deltarow}$\uparrow 19.0\%$ & \cellcolor{deltarow} & \cellcolor{deltarow}$\uparrow 21.0\%$ & \cellcolor{deltarow} & \cellcolor{deltarow}$\uparrow 21.2\%$ & \cellcolor{deltarow} & \cellcolor{deltarow}$\uparrow 3.4\%$ & \cellcolor{deltarow} & \cellcolor{deltarow}$\uparrow 7.8\%$ \\
\midrule

\multirow{5}{*}{\makecell{\textbf{DS-32B} \\ \textbf{(bs=1)}}}
& AR         & 50.72 & 1.00$\times$ & 50.90 & 1.00$\times$ & 50.83 & 1.00$\times$ & 50.96 & 1.00$\times$ & 48.82 & 1.00$\times$ \\
& PEARL      & 90.00 & 1.77$\times$ & 86.22 & 1.69$\times$ & 94.97 & 1.87$\times$ & 86.32 & 1.69$\times$ & 99.72 & 2.04$\times$ \\
& Standalone & \underline{118.55} & \underline{2.34$\times$} & \underline{114.83} & \underline{2.26$\times$} & \underline{115.44} & \underline{2.27$\times$} & \underline{123.26} & \underline{2.42$\times$} & \underline{115.58} & \underline{2.37$\times$} \\
& \cellcolor{spectrerow}\textbf{SPECTRE} & \cellcolor{spectrerow}\textbf{126.97} & \cellcolor{spectrerow}\textbf{2.50$\times$} & \cellcolor{spectrerow}\textbf{120.07} & \cellcolor{spectrerow}\textbf{2.36$\times$} & \cellcolor{spectrerow}\textbf{120.48} & \cellcolor{spectrerow}\textbf{2.37$\times$} & \cellcolor{spectrerow}\textbf{140.60} & \cellcolor{spectrerow}\textbf{2.76$\times$} & \cellcolor{spectrerow}\textbf{147.85} & \cellcolor{spectrerow}\textbf{3.03$\times$} \\
& \cellcolor{deltarow}$\Delta$ & \cellcolor{deltarow} & \cellcolor{deltarow}$\uparrow 6.8\%$ & \cellcolor{deltarow} & \cellcolor{deltarow}$\uparrow 4.4\%$ & \cellcolor{deltarow} & \cellcolor{deltarow}$\uparrow 4.4\%$ & \cellcolor{deltarow} & \cellcolor{deltarow}$\uparrow 14.1\%$ & \cellcolor{deltarow} & \cellcolor{deltarow}$\uparrow 27.8\%$ \\
\midrule

\multirow{5}{*}{\makecell{\textbf{DS-32B} \\ \textbf{(bs=16)}}}
& AR         & 682.53 & 1.00$\times$ & 681.51 & 1.00$\times$ & 681.78 & 1.00$\times$ & 547.01 & 1.00$\times$ & 428.59 & 1.00$\times$ \\
& PEARL      & 936.72 & 1.37$\times$ & 933.64 & 1.37$\times$ & 901.64 & 1.32$\times$ & 633.15 & 1.16$\times$ & 608.60 & 1.42$\times$ \\
& Standalone & \underline{1292.60} & \underline{1.89$\times$} & \underline{1355.10} & \underline{1.99$\times$} & \underline{1261.70} & \underline{1.85$\times$} & \underline{915.66} & \underline{1.67$\times$} & \underline{637.10} & \underline{1.49$\times$} \\
& \cellcolor{spectrerow}\textbf{SPECTRE} & \cellcolor{spectrerow}\textbf{1324.66} & \cellcolor{spectrerow}\textbf{1.94$\times$} & \cellcolor{spectrerow}\textbf{1460.37} & \cellcolor{spectrerow}\textbf{2.14$\times$} & \cellcolor{spectrerow}\textbf{1351.04} & \cellcolor{spectrerow}\textbf{1.98$\times$} & \cellcolor{spectrerow}\textbf{978.20} & \cellcolor{spectrerow}\textbf{1.79$\times$} & \cellcolor{spectrerow}\textbf{787.07} & \cellcolor{spectrerow}\textbf{1.84$\times$} \\
& \cellcolor{deltarow}$\Delta$ & \cellcolor{deltarow} & \cellcolor{deltarow}$\uparrow 2.6\%$ & \cellcolor{deltarow} & \cellcolor{deltarow}$\uparrow 7.5\%$ & \cellcolor{deltarow} & \cellcolor{deltarow}$\uparrow 7.0\%$ & \cellcolor{deltarow} & \cellcolor{deltarow}$\uparrow 7.2\%$ & \cellcolor{deltarow} & \cellcolor{deltarow}$\uparrow 23.5\%$ \\
\midrule

\multirow{5}{*}{\makecell{\textbf{DS-32B} \\ \textbf{(bs=32)}}}
& AR         & 1156.82 & 1.00$\times$ & 1149.42 & 1.00$\times$ & 1143.46 & 1.00$\times$ & 853.33 & 1.00$\times$ & 565.17 & 1.00$\times$ \\
& PEARL      & 1023.68 & 0.88$\times$ & 1150.46 & 1.00$\times$ & 1101.21 & 0.96$\times$ & 848.33 & 0.99$\times$ & \underline{685.79} & \underline{1.21$\times$} \\
& Standalone & \underline{2081.00} & \underline{1.80$\times$} & \underline{1973.52} & \underline{1.72$\times$} & \underline{2049.75} & \underline{1.79$\times$} & \underline{933.02} & \underline{1.09$\times$} & 622.29 & 1.10$\times$ \\
& \cellcolor{spectrerow}\textbf{SPECTRE} & \cellcolor{spectrerow}\textbf{2148.10} & \cellcolor{spectrerow}\textbf{1.86$\times$} & \cellcolor{spectrerow}\textbf{2215.11} & \cellcolor{spectrerow}\textbf{1.93$\times$} & \cellcolor{spectrerow}\textbf{2101.60} & \cellcolor{spectrerow}\textbf{1.84$\times$} & \cellcolor{spectrerow}\textbf{1397.53} & \cellcolor{spectrerow}\textbf{1.64$\times$} & \cellcolor{spectrerow}\textbf{920.98} & \cellcolor{spectrerow}\textbf{1.63$\times$} \\
& \cellcolor{deltarow}$\Delta$ & \cellcolor{deltarow} & \cellcolor{deltarow}$\uparrow 3.3\%$ & \cellcolor{deltarow} & \cellcolor{deltarow}$\uparrow 12.2\%$ & \cellcolor{deltarow} & \cellcolor{deltarow}$\uparrow 2.5\%$ & \cellcolor{deltarow} & \cellcolor{deltarow}$\uparrow 50.5\%$ & \cellcolor{deltarow} & \cellcolor{deltarow}$\uparrow 34.1\%$ \\
\midrule

\multirow{5}{*}{\makecell{\textbf{DS-32B} \\ \textbf{(bs=64)}}}
& AR         & 1697.20 & 1.00$\times$ & 1693.34 & 1.00$\times$ & 1681.12 & 1.00$\times$ & 924.19 & 1.00$\times$ & 587.54 & 1.00$\times$ \\
& PEARL      & 969.03 & 0.57$\times$ & 908.49 & 0.54$\times$ & 1023.42 & 0.61$\times$ & 946.97 & 1.02$\times$ & \underline{719.87} & \underline{1.23$\times$} \\
& Standalone & \underline{2637.40} & \underline{1.55$\times$} & \underline{2697.26} & \underline{1.59$\times$} & \underline{2494.65} & \underline{1.48$\times$} & \underline{968.29} & \underline{1.05$\times$} & 657.09 & 1.12$\times$ \\
& \cellcolor{spectrerow}\textbf{SPECTRE} & \cellcolor{spectrerow}\textbf{2819.25} & \cellcolor{spectrerow}\textbf{1.66$\times$} & \cellcolor{spectrerow}\textbf{2875.32} & \cellcolor{spectrerow}\textbf{1.70$\times$} & \cellcolor{spectrerow}\textbf{2708.55} & \cellcolor{spectrerow}\textbf{1.61$\times$} & \cellcolor{spectrerow}\textbf{1446.21} & \cellcolor{spectrerow}\textbf{1.56$\times$} & \cellcolor{spectrerow}\textbf{966.69} & \cellcolor{spectrerow}\textbf{1.65$\times$} \\
& \cellcolor{deltarow}$\Delta$ & \cellcolor{deltarow} & \cellcolor{deltarow}$\uparrow 7.1\%$ & \cellcolor{deltarow} & \cellcolor{deltarow}$\uparrow 6.9\%$ & \cellcolor{deltarow} & \cellcolor{deltarow}$\uparrow 8.8\%$ & \cellcolor{deltarow} & \cellcolor{deltarow}$\uparrow 48.6\%$ & \cellcolor{deltarow} & \cellcolor{deltarow}$\uparrow 34.1\%$ \\
\midrule

\multirow{5}{*}{\makecell{\textbf{DS-32B} \\ \textbf{(bs=128)}}}
& AR         & 2195.28 & 1.00$\times$ & 2200.47 & 1.00$\times$ & 2198.42 & 1.00$\times$ & 903.00 & 1.00$\times$ & 582.58 & 1.00$\times$ \\
& PEARL      & 637.58 & 0.29$\times$ & 637.59 & 0.29$\times$ & 768.77 & 0.35$\times$ & 987.80 & 1.09$\times$ & \underline{683.28} & \underline{1.17$\times$} \\
& Standalone & \underline{2734.99} & \underline{1.25$\times$} & \underline{2818.41} & \underline{1.28$\times$} & \underline{2595.55} & \underline{1.18$\times$} & \underline{996.86} & \underline{1.10$\times$} & 655.23 & 1.12$\times$ \\
& \cellcolor{spectrerow}\textbf{SPECTRE} & \cellcolor{spectrerow}\textbf{3199.90} & \cellcolor{spectrerow}\textbf{1.46$\times$} & \cellcolor{spectrerow}\textbf{3249.42} & \cellcolor{spectrerow}\textbf{1.48$\times$} & \cellcolor{spectrerow}\textbf{2969.55} & \cellcolor{spectrerow}\textbf{1.35$\times$} & \cellcolor{spectrerow}\textbf{1497.50} & \cellcolor{spectrerow}\textbf{1.66$\times$} & \cellcolor{spectrerow}\textbf{967.50} & \cellcolor{spectrerow}\textbf{1.66$\times$} \\
& \cellcolor{deltarow}$\Delta$ & \cellcolor{deltarow} & \cellcolor{deltarow}$\uparrow 16.8\%$ & \cellcolor{deltarow} & \cellcolor{deltarow}$\uparrow 15.6\%$ & \cellcolor{deltarow} & \cellcolor{deltarow}$\uparrow 14.4\%$ & \cellcolor{deltarow} & \cellcolor{deltarow}$\uparrow 50.9\%$ & \cellcolor{deltarow} & \cellcolor{deltarow}$\uparrow 41.9\%$ \\
\bottomrule
\end{tabular}
}
\caption{
Throughput results at temperature 0.6 for Qwen3-32B and DeepSeek-R1-Distill-Qwen-32B across all batch sizes.
For each benchmark, the SOTA result is underlined.
$\Delta$ reports the relative improvement of SPECTRE over the SOTA method.
}
\label{tab:temp06_all_bs_results}
\end{table*}

%% file: tex/A2_experiment_detail.tex
\section{Experiment Settings}
\label{app:Experiment Settings}

\subsection{Detailed Experimental Setup for Table.\ref{tab:main_compact_results},\ref{tab:appendix_small_bs},\ref{tab:accept_len_avg}}
\label{app:Detailed_Experimental_Setup_for_Table}
\paragraph{Workload generation.}
To emulate realistic online serving traffic, we generate request arrivals according to a Poisson process. Specifically, the inter-arrival time between two consecutive requests is sampled as
\begin{equation}
\Delta t \sim \mathrm{Exp}(1/\mathrm{QPS}),
\end{equation}
where $\mathrm{QPS}$ denotes the target request rate. This setup approximates bursty and stochastic production workloads.

\paragraph{Batch size control.}
We control the effective batch size through the maximum request concurrency parameter (\texttt{max-concurrency}). To efficiently obtain stable measurements across different regimes, we jointly adjust QPS and the number of evaluation samples for each batch size. Concretely, we use
\begin{itemize}
    \item Batch sizes: $\{1,16,32,64,128\}$,
    \item QPS: $\{1,4,8,16,32\}$,
    \item Number of samples: $\{8,64,128,256,512\}$,
\end{itemize}
respectively. Larger batch sizes correspond to higher request rates and more samples to ensure sufficient system saturation and statistical stability.

\paragraph{Decoding configuration.}
We implement SPECTRE in \texttt{SGLang v0.5.7}. Speculative decoding is a lossless acceleration technique; therefore, we focus on throughput rather than output quality. We fix the output length of each request to 1024 tokens. To avoid early termination due to end-of-sequence tokens, we enable \texttt{ignore\_eos} during decoding. This ensures that all methods are evaluated under identical decoding lengths and eliminates variance caused by early stopping.

\paragraph{Dataset-specific setup.}
For mathematical reasoning datasets (GSM8K, Math500, and Minerva Math), we construct few-shot prompts by prepending the solutions of the first five problems as demonstrations. This provides consistent reasoning context across all methods. 
For ShareGPT, we follow the official \texttt{sglang} benchmarking pipeline and use \texttt{bench\_serving.py} to generate realistic conversational workloads. The input prompt length is fixed to 4000 tokens to simulate long-context scenarios.

\paragraph{Model parallelism.}
In these experiments, both the target and draft models are deployed with TP size 1. As a result, there is no significant compute imbalance between the two models. Consequently, we do not apply prompt compression in this setting, allowing us to isolate the effect of the speculative decoding strategy itself.

\subsection{Detailed experimental setup for Qwen3-235B-A22B}
\label{app:Experimental setup for Qwen3-235B-A22B}
We evaluate SPECTRE on a large-model setting using the Qwen3-235B-A22B / Qwen3-0.6B pair, as shown in Fig.~\ref{fig:qwen3-235b}. 
In this configuration, the target model is deployed with TP size 8, while the draft model uses TP=1, reflecting the realistic compute imbalance between large and small models in production systems.

To emulate online serving conditions, we set the request rate to $\mathrm{QPS}=32$ and the maximum request concurrency to 128. 
Each request generates 1024 output tokens, and \texttt{ignore\_eos} is enabled to ensure consistent decoding length across all methods.

Due to the substantial compute gap between the target and draft models, we apply draft-side context compression to reduce latency. 
Specifically, given a prompt of length $S$, we retain only a fraction $p$ of tokens by keeping the first $pS/2$ tokens and the last $pS/2$ tokens, with $p=0.1$ in our experiments. 
This reduces the draft-side prefill and decoding cost, enabling better overlap between draft generation and target-side verification.

\subsection{Mixed Draft Traffic Setup}
\label{app:mixed_traffic}

We evaluate system robustness under mixed draft-side traffic by fixing the target workload and varying the background load on the draft server.
The target request rate is fixed, with a maximum request concurrency of 128. 
On the draft side, we inject background user traffic with request rates of 1, 2, 4, and 8 QPS. 
For each draft QPS setting, a total of 64 requests are issued, while the maximum request concurrency of the draft server is fixed at 256. 
For each background load level, we first measure the draft throughput, and then evaluate the resulting target throughput under the same conditions.

\subsection{System Benefit Evaluation Setup}
\label{app:benefit_setup}

We evaluate system-level benefit efficiency in terms of dollar/1000s/GPU, which measures the revenue generated per unit GPU time by jointly accounting for both target and draft serving. The target workload is fixed to 32 QPS with a maximum request concurrency of 128. For SPECTRE and PEARL, the draft model is deployed on a separate GPU (TP1), while for other baselines (e.g., EAGLE3 and Standalone), the draft and target models are co-located on the same GPU. All models use TP1 unless otherwise specified, except Qwen3-235B-A22B, which runs with TP8.
To compute the system-level benefit, we explicitly account for both throughput and resource cost. For Qwen3-32B and DeepSeek-R1-Distill-Qwen-32B, the target model is deployed with TP1 and paired with a smaller draft model (Qwen3-0.6B or DeepSeek-R1-Distill-Qwen-1.5B) on a separate GPU (also TP1), resulting in a total cost of two GPUs. For Qwen3-235B-A22B, the target model is deployed across 8 GPUs (TP8), together with a draft model on an additional GPU (TP1), leading to a total cost of nine GPUs. For AR, EAGLE3, Standalone, and SPECTRE, we define the benefit as the total served tokens multiplied by token price, i.e., $(\text{target throughput} + \text{draft throughput}) \times \text{price}$, normalized by the number of GPUs. In contrast, for \textsc{PEARL}, since the draft model does not serve regular user traffic, its contribution to revenue is excluded, and the benefit is computed solely based on target throughput divided by the total GPU count.
For the Qwen3-235B-A22B (TP=8) / Qwen3-0.6B (TP=1) configuration, the target-side compute is substantially stronger, making the draft side the bottleneck; accordingly, we set a lower background load on the draft server (4 QPS with maximum concurrency 100). For other settings, we use a higher background draft load (8 QPS with maximum concurrency 256).
Token pricing follows public API rates\footnote{https://help.aliyun.com/zh/model-studio/model-pricing} (as of 04/23/2026): 
Qwen3-32B (\$3/M), Qwen3-0.6B (\$0.45/M), Qwen3-235B-A22B (\$3/M), 
DeepSeek-R1-Distill-Qwen-32B (\$0.9/M), and DeepSeek-R1-Distill-Qwen-1.5B (free).